%% file: main.tex
\documentclass[aps,physrev,reprint,superscriptaddress,nofootinbib,floatfix,longbibliography]{revtex4-2}

\usepackage[T1]{fontenc}
\usepackage[utf8]{inputenc}
\usepackage{lmodern}
\usepackage{amsmath,amssymb,bm,mathtools}
\usepackage{graphicx}
\usepackage{xcolor}
\usepackage{etoolbox}
\usepackage{booktabs}
\usepackage{xurl}
\usepackage{microtype}
\usepackage[colorlinks=true,linkcolor=blue,citecolor=blue,urlcolor=blue]{hyperref}
\hypersetup{
 pdftitle={Fisher-information retention under local driving in non-Hermitian feed-forward chains},
 pdfauthor={Qingrui Bai, Zhichao Li, and Junlong Kou},
 pdfsubject={Local-drive constraints and Fisher-information retention in non-Hermitian sensors}
}
\DeclareMathOperator{\Tr}{Tr}
\DeclareMathOperator{\rank}{rank}
\DeclareMathOperator{\ran}{ran}
\DeclareMathOperator{\diag}{diag}

\newcommand{\ii}{\mathrm{i}}
\newcommand{\norm}[1]{\left\lVert#1\right\rVert}

\newcommand{\Ccal}{\mathcal C}
\newcommand{\Ecal}{\mathcal E}

\newcommand{\Gcal}{\mathcal G}

\newcommand{\Rcal}{\mathcal R}
\InputIfFileExists{generated_numbers.tex}{}{}

\begin{document}

\title{Fisher-information retention under local driving in non-Hermitian feed-forward chains}

\author{Qingrui Bai}
\affiliation{School of Electronic Science and Engineering, Nanjing University, Nanjing 210023, China}
\affiliation{School of Integrated Circuits, Nanjing University, Suzhou 215163, China}
\affiliation{National Key Laboratory of Transient Impact, Nanjing 210023, China}
\author{Zhichao Li}
\affiliation{School of Electronic Science and Engineering, Nanjing University, Nanjing 210023, China}
\affiliation{School of Integrated Circuits, Nanjing University, Suzhou 215163, China}
\affiliation{National Key Laboratory of Transient Impact, Nanjing 210023, China}
\author{Junlong Kou}
\email[Contact author: ]{jlkou@nju.edu.cn}
\affiliation{School of Electronic Science and Engineering, Nanjing University, Nanjing 210023, China}
\affiliation{School of Integrated Circuits, Nanjing University, Suzhou 215163, China}
\affiliation{National Key Laboratory of Transient Impact, Nanjing 210023, China}
\affiliation{Key Laboratory of Intelligent Optical Sensing and Manipulation, Ministry of Education, Nanjing University, Nanjing 210093, China}
\affiliation{Jiangsu Key Laboratory of Semiconductor Laser and Sensing Technology, Suzhou 215163, China}
\affiliation{Wujin-NJU Institute of Future Technology, Changzhou 213153, China}

\begin{abstract}
When local drive hardware is installed before a sensing task is revealed,
directional non-Hermitian buildup creates a tradeoff between targeted
preparation and spatial coverage.  We quantify this tradeoff for stable linear
scatterers with proper complex Gaussian output statistics.  For parameters
encoded in the internal generator, and at fixed coherent internal resource,
the drive hardware enters the optimized mean-response Fisher information only
through the projector onto the internally reachable subspace, while the
preparation Gramian sets the source cost.  In an
on-resonance feed-forward chain, nested Green-function columns decompose into
disjoint port intervals, and the site-local retention factors in each interval
sum exactly to unity.  Directionality concentrates this unit downstream, so
the worst-site retention decays exponentially when one port serves many sites;
for \(m\) equal intervals, a single probe shared by all tasks has a worst-site
retention lower by a further factor \(m\) than task-specific reoptimization.
Joint source-power and
internal-resource constraints connect this fixed-resource law to
source-limited preparation.
A noisy RLC/VCCS model illustrates the retention law with stored
electromagnetic energy, covariance Fisher information, and component
disorder.  The resulting bounds turn directional buildup into a port-layout
criterion: source savings for a selected task are accompanied by a spatial
coverage cost for tasks unknown at the hardware-design stage.
\end{abstract}

\maketitle

\section{Introduction}

Gain, loss, exceptional points (EPs), and the non-Hermitian skin effect
(NHSE) reshape eigenvalue splittings, steady-state fields, and directional
Green functions~\cite{ElGanainy2018,MiriAlu2019,Ashida2020,Bergholtz2021}.
For sensing, these responses acquire operational meaning only through the
statistics of the measured record.  Input--output analyses have shown that EP
splitting or internal buildup alone does not determine precision because
output coupling and noise enter the same measurement problem
~\cite{Langbein2018,LauClerk2018,McDonaldClerk2020,LoughlinSudhir2024}.

Optimized drives and nonreciprocal propagation can enhance a specified task
under source- or flux-normalized signal-to-noise and Fisher metrics
~\cite{Bao2022,BlanchardMcDonaldStJean2025,WiersigRotter2026}.
Maximum-information states optimize the incident wavefront
~\cite{Bouchet2021,Verreussel2026}, while continuous-Gaussian metrology
distinguishes global information from information carried into the monitored
environment~\cite{YokomizoClerkAshida2026}.  A complementary spatial
description assigns densities and fluxes to Fisher information as it propagates
through a wave system~\cite{Hupfl2024,Weimar2025}.  Here the spatial question
arises at the drive: how well can a fixed collection of local ports cover tasks
whose locations are specified only after installation?

\emph{Operational protocol.} The drive matrix \(\bm B\) is installed before
the task location is known.  The physical-port and ideal full-state benchmarks
use the same generator, internal metric, receiver, output covariance, and
record count; only the drive matrix changes, with
\(\bm B_{\rm full}=\bm I\).  After a task is selected, the task-specific
protocol reoptimizes the complex drive amplitudes, whereas the common-probe
protocol uses one drive for all tasks.  Retention is the resulting ratio of
optimized mean Fisher informations.  This protocol separates drive-side access
from receive-side combining and finite-time state-transfer
cost~\cite{ZhangMiLAC2026,KlicksteinSorrentino2020Gramian}.

This ordering occurs naturally in spatially distributed photonic, microwave,
acoustic, and mechanical sensors: coupling ports are set during fabrication,
whereas the perturbation can appear at different locations after deployment.
Fisher retention measures the information loss imposed by that fixed access
geometry.

The central result is an exact interval conservation law for a strictly
feed-forward chain.  Nested Green-function columns split the chain into
nonoverlapping port intervals within which the site-local retention factors sum
to unity.  This law fixes the exponential worst-task penalty, the port density
needed to maintain a prescribed information floor, and the additional cost of
a common probe.  The reachable-subspace projector gives the corresponding
fixed-resource result for a general linear scatterer, while the preparation
Gramian and a joint resource constraint restore source cost.  A noisy
reduced-order circuit illustrates how covariance information changes the total
Fisher ratio; reverse hopping and pole-identical dimers provide complementary
tests beyond the ideal chain.  The resulting bounds directly specify the port
density and placement needed to maintain a desired worst-task information
level.

\section{Model and Gaussian Fisher information}

At a fixed real frequency $\omega$, consider
\begin{equation}
 \bm b=(\bm D-\ii\bm C\bm G\bm B)\bm s+\bm n.
 \label{eq:io}
\end{equation}
Here \(\bm G=(\bm H_{\rm eff}-\omega\bm I)^{-1}\) and
\(\bm S\equiv\bm D-\ii\bm C\bm G\bm B\).
The vectors $\bm s\in\mathbb C^p$ and $\bm b\in\mathbb C^q$ are the
coherent input and monitored output.  The noise $\bm n$ is proper complex
Gaussian with covariance
$\bm\Sigma_c(\bm\theta,\omega)\succ0$.  The operating point is stable and
the resolvent exists.  The parameter vector $\bm\theta$ may contain the
target parameter and calibration or environmental nuisances.

We use the proper complex convention
\(\mathbb E[\bm n\bm n^\dagger]=\bm\Sigma_c\) and
\(\mathbb E[\bm n\bm n^{\mathsf T}]=0\).  For one statistically independent
complex record, let \(\bm d_\alpha=\partial_\alpha\bm\mu\),
\(\bm\Sigma_\alpha=\partial_\alpha\bm\Sigma_c\), and
\(\bm\mu=\bm S\bm s\).  The Fisher matrix for real parameters is
~\cite{SchreierScharf2010,Kay1993}
\begin{equation}
 F_{\alpha\beta}=2\operatorname{Re}(\bm d_\alpha^\dagger\bm\Sigma_c^{-1}\bm d_\beta)+\Tr(\bm\Sigma_c^{-1}\bm\Sigma_\alpha\bm\Sigma_c^{-1}\bm\Sigma_\beta).
 \label{eq:Ffull}
\end{equation}
We write
\begin{equation}
 \bm T_\alpha=\bm\Sigma_c^{-1/2}\partial_\alpha\bm S
 \label{eq:tangent}
\end{equation}
for the noise-whitened scattering tangent.  We call the two terms in
Eq.~\eqref{eq:Ffull} the mean FI \(F_{\mu,\alpha\beta}\) and covariance FI
\(F_{\Sigma,\alpha\beta}\).  For a scalar target,
\(F_\Sigma=\Tr[(\bm\Sigma_c^{-1}\partial_\theta\bm\Sigma_c)^2]\).
The mean FI depends on the coherent probe, whereas the covariance FI is set by
\(\bm\Sigma_c\) and its parameter derivative.  Under the operational protocol,
the projector law below governs the mean-response ratio; covariance information
enters through the second term of Eq.~\eqref{eq:Ffull}.

For independent records, coherent probes enter the mean FI through
\(\bm\Xi=\sum_k n_k\bm s_k\bm s_k^\dagger\).  Covariance information scales
with the record count.  Multiprobe and nuisance-adjusted
extensions are derived in Secs.~S1 and S6 of the Supplemental
Material.  Related experimental-design and
nuisance-parameter formulations appear in
Refs.~\cite{ChalonerVerdinelli1995,VanTrees2001,Naikoo2023Multiparameter}.

\section{Fixed-resource reachability and source price}

\begin{figure}[!ht]
 \centering
 \includegraphics[width=0.98\columnwidth]{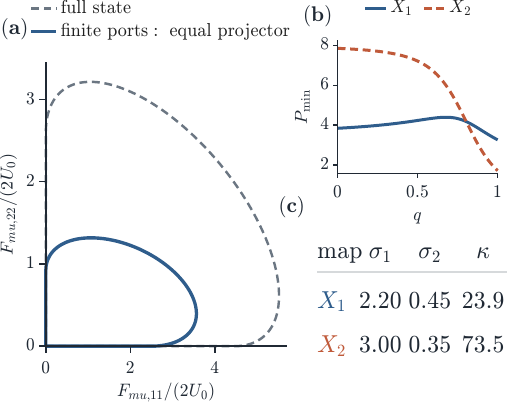}
\caption{\textbf{Reachable Fisher region and preparation cost.}
 (a) In a stable four-mode example, two rank-2 preparation maps with the same
 reachable projector generate the same
 \((F_{\mu,11},F_{\mu,22})/(2U_0)\) boundary (blue); the ideal full-state drive benchmark
 gives the gray dashed boundary.  This convex multiprobe region is generated
 by \(\bm\Xi_z\succeq0\) and \(\Tr\bm\Xi_z\le U_0\).
 (b) The maps require different minimum source powers along the common
 boundary, parametrized by the task weight \(q\).
 (c) Their nonzero singular values and preparation-Gramian condition numbers
 quantify the difference in source cost.  Here \(\bm W_0=\bm X\bm X^\dagger\).}
 \label{fig:region}
\end{figure}

Two drive configurations can span the same internal subspace while requiring
different source powers to prepare a field within it.  The four-mode example in
Fig.~\ref{fig:region} makes this distinction explicit.  Its fixed-\(U\) Fisher
boundaries coincide [panel (a)], but their minimum source costs differ along
that boundary [panel (b)] because their preparation spectra differ [panel
(c)].  Section~S2 of the Supplemental Material
 extends the result
for a general positive-semidefinite source metric.

\subsection{Reachable mean-FI region}

Let the internal state be $\bm a=\bm G\bm B\bm s$ and define an internal quadratic resource
\begin{equation}
 U=\bm a^\dagger\bm Q\bm a,\qquad \bm Q\succ0.
 \label{eq:energy}
\end{equation}
In the RLC example, \(U\) is the stored electromagnetic energy of the coherent
mean field.  Section~S2 of the
Supplemental Material treats parameter-dependent metrics and external
couplings.
Define
\begin{equation}
 \bm X=\bm Q^{1/2}\bm G\bm B,\qquad
 \bm\Pi_X=\bm X\bm X^+,
 \label{eq:XPi}
\end{equation}
For the fixed-resource factorization below, parameters are encoded through
\(\bm H_{\rm eff}\), while \(\bm B\), \(\bm C\), and \(\bm D\) are parameter
independent at the operating point.  For each such parameter, define
\begin{equation}
 \bm V_\alpha=\partial_\alpha\bm H_{\rm eff},\qquad
 \bm A_\alpha=
 \bm\Sigma_c^{-1/2}\bm C\bm G\bm V_\alpha\bm Q^{-1/2}.
 \label{eq:parameterMaps}
\end{equation}
For a coherent ensemble, let
\begin{equation}
 \bm\Xi_z=\sum_k n_k\bm z_k\bm z_k^\dagger,\qquad
 \bm z_k=\bm X\bm s_k.
 \label{eq:internalMoment}
\end{equation}
Any admissible internal moment has rank no larger than \(\rank\bm X\) and can
therefore be realized using at most \(\rank\bm X\) independently weighted
coherent settings.  The attainable set of mean-FI matrices under
the aggregate internal-resource budget \(\Tr\bm\Xi_z\le U_0\) is
\begin{equation}
 \begin{split}
 \mathfrak R_{\mu,U}={}\Big\{&
 [\,2\operatorname{Re}\Tr(
 \bm\Xi_z\bm A_\alpha^\dagger\bm A_\beta)\,]_{\alpha\beta}:\\[-2pt]
 &\bm\Xi_z\succeq0,\quad
 \bm\Pi_X\bm\Xi_z\bm\Pi_X=\bm\Xi_z,\quad
 \Tr\bm\Xi_z\le U_0\Big\}.
 \end{split}
 \label{eq:FisherRegion}
\end{equation}
Every admissible \(\bm\Xi_z\) is then realizable.
Preparation maps with the same
\(\bm\Pi_X\), internal metric, and downstream maps generate the same region.
Reachable-subspace inclusion likewise implies region inclusion.  The scalar
optimum below is rank one and requires one coherent setting.  Sections~S1 and
S6 of the Supplemental Material give the record-limited and nuisance-adjusted
forms.

For a scalar target, set $\bm V=\bm V_\theta$ and $\bm A=\bm A_\theta$.  Maximizing a linear functional over Eq.~\eqref{eq:FisherRegion} gives
\begin{equation}
 F_{\mu,U}^{\star}
 =2U_0\norm{\bm A\bm\Pi_X}_2^2.
 \label{eq:storedTheorem}
\end{equation}
For fixed \(\bm Q\) and downstream maps \(\{\bm A_\alpha\}\), the drive-hardware
dependence of Eqs.~\eqref{eq:FisherRegion} and~\eqref{eq:storedTheorem} is
completely specified by \(\bm\Pi_X\).  This endpoint still has a source cost,
set by the
nonzero singular values of the source-whitened preparation map.

For $\bm R\succ0$, let
\(\bm Y=\bm X\bm R^{-1/2}\) and
\(\bm W_0=\bm Y\bm Y^\dagger\).  The minimum aggregate source-resource cost of an admissible internal moment is
\begin{equation}
 P_{\min}(\bm\Xi_z)=\Tr(\bm W_0^+\bm\Xi_z).
 \label{eq:sourcePrice}
\end{equation}
This is the minimum weighted incident power (or squared input norm when
\(\bm R=\bm I\)) required to prepare that internal second moment.  If several
moments yield the same Fisher matrix, its preparation cost is the minimum of
Eq.~\eqref{eq:sourcePrice} over that set.
The feed-forward chain now turns this separation between accessible fields and
their source price into an exact spatial conservation law.

\section{Local-port law for Fisher retention}

For the generator-encoded target introduced above,
\begin{equation}
 \partial_\theta\bm S=\ii\bm C\bm G\bm V\bm G\bm B.
 \label{eq:dSfactor}
\end{equation}
The Green factor adjacent to \(\bm B\) prepares the internal field; the factor
adjacent to \(\bm C\) propagates the parameter-induced field to the monitored
outputs.  Fixing the internal resource
removes the magnitudes of the nonzero preparation singular values from the
preparation factor.  Downstream propagation, parameter action, and monitored emission remain
in \(\bm A_\alpha\)~\cite{WiersigRotter2026}.

\subsection{Task-specific retention}

For a rank-one parameter channel \(\bm V=\bm v\bm w^\dagger\), set
\(\widetilde{\bm w}=\bm Q^{-1/2}\bm w\) and introduce
\begin{align}
 \Ecal&=\norm{\bm\Sigma_c^{-1/2}\bm C\bm G\bm v}^2,\nonumber\\
 \alpha&=\frac{\widetilde{\bm w}^\dagger\bm\Pi_X\widetilde{\bm w}}
 {\norm{\widetilde{\bm w}}^2},\nonumber\\
 \beta&=\frac{\norm{\bm X^\dagger\widetilde{\bm w}}^2}
 {\widetilde{\bm w}^\dagger\bm\Pi_X\widetilde{\bm w}}.
 \label{eq:Eab}
\end{align}
For $\alpha>0$ and the Euclidean source metric $\bm R=\bm I$, one then has
\begin{align}
 F_{\mu,P}^{\star}
 &=2P_0\,\Ecal\,\norm{\widetilde{\bm w}}^2\alpha\beta,
 \label{eq:factorP}\\
 F_{\mu,U}^{\star}
 &=2U_0\,\Ecal\,\norm{\widetilde{\bm w}}^2\alpha.
 \label{eq:factorU}
\end{align}
Here \(\beta\) measures preparation buildup, \(\alpha\) is the normalized
squared projection of the task onto the reachable subspace, and \(\Ecal\)
measures emission after noise whitening.  At fixed $P$, all three factors
enter.  At fixed $U$, \(\beta\) drops out.  Singular source metrics
 and the case \(\alpha=0\) are treated in Sec.~S2 of the Supplemental
Material.

Covariance information follows from the second term of Eq.~\eqref{eq:Ffull}.
We use \(\bm Q=\bm I\) for equal modal occupation in the chain and the
electromagnetic stored-energy metric in the circuit.  Coupled-mode and
dwell-time metrics are discussed in Sec.~S2 of the
Supplemental Material
~\cite{Wigner1955,Smith1960,MahauxWeidenmuller1969,WiersigRotter2026,
YokomizoClerkAshida2026}.

\subsection{Feed-forward chain}

Let
$\bm T_+=\sum_{j=1}^{N-1}|j+1\rangle\langle j|$ denote the forward shift and consider the stable order-$N$ feed-forward chain
\begin{equation}
 \bm H_J=-\ii\gamma\bm I+\kappa\bm T_+,
 \qquad \rho=\kappa/\gamma,
 \label{eq:Jordan}
\end{equation}
whose poles all remain at $-\ii\gamma$.  At \(\omega=0\), take \(\gamma>0\),
\(\kappa\ge0\), \(\bm Q=\bm I\), positive budgets \(P_0,U_0\), a source at
site \(1\), and a rank-one end task
\(\bm V_N=\bm v_N\bm e_N^\dagger\).  With
\(\Ecal_N=\norm{\bm\Sigma_c^{-1/2}\bm C\bm G\bm v_N}^2>0\), the
emission-normalized, per-unit-source coefficient
\begin{equation}
 \mathcal I_{P,N}\equiv
 \frac{F_{\mu,P,N}^{\star}}{2P_0\Ecal_N}
 =|\bm e_N^\dagger\bm G\bm e_1|^2
 \label{eq:JordanFisherRate}
\end{equation}
then obeys
\begin{equation}
 \gamma^2\mathcal I_{P,N}=\rho^{2(N-1)}.
 \label{eq:JordanSource}
\end{equation}
At fixed $U$, the corresponding retention is
\begin{equation}
 \alpha_N\equiv
 \frac{F_{\mu,U,N}^{\star}}{2U_0\Ecal_N}
 =\frac{\rho^{2(N-1)}}{\sum_{r=0}^{N-1}\rho^{2r}}\le1.
 \label{eq:JordanEnergy}
\end{equation}
Writing \(S_N=\sum_{r=0}^{N-1}\rho^{2r}\), the dimensionless one-port
preparation buildup is
\(\Gcal_N\equiv\gamma^2\norm{\bm G\bm e_1}^2=S_N\).  For \(\rho\ge1\),
the weakest local task is the upstream site, with
\(\Ccal_{\min}=1/S_N\), and therefore
\begin{equation}
 \Gcal_N\Ccal_{\min}=1.
 \label{eq:gainRetentionReciprocity}
\end{equation}
In this uniform one-port chain, the buildup that reduces the source cost of a
downstream task is exactly reciprocal to the retained worst-site coverage at
fixed internal resource.
For \(\rho>1\), \(\mathcal I_{P,N}\) grows exponentially although every pole
remains fixed at \(-\ii\gamma\).  At fixed internal resource, the corresponding
normalized quantity instead saturates at
\(\alpha_N\to1-\rho^{-2}\).  The mean FI per unit source is
\(F_{\mu,P,N}^{\star}/P_0=2\Ecal_N\mathcal I_{P,N}\), so its size
dependence also contains the emission factor \(\Ecal_N\).
Equations~\eqref{eq:JordanSource} and~\eqref{eq:JordanEnergy} expose the
opposite system-size scaling of the two resource normalizations; additional
\(\rho\) regimes are summarized in Sec.~S4 of the Supplemental Material.

For \(U_0>0\), let
\(\{\widetilde{\bm w}_j\}_{j=1}^{N}\) be a complete orthonormal basis of the
internal task space, with a nonzero full-state-drive Fisher denominator for every
task.  The retained mean FI is
\begin{equation}
 \Ccal_j=
 \frac{F_{\mu,U,B,j}^{\star}}{F_{\mu,U,\rm full,j}^{\star}}
 =
 \frac{\widetilde{\bm w}_j^\dagger\bm\Pi_X\widetilde{\bm w}_j}
 {\norm{\widetilde{\bm w}_j}^2}.
 \label{eq:coverage}
\end{equation}
It obeys
\begin{equation}
 0\le\Ccal_j\le1,\qquad
 \sum_{j=1}^N\Ccal_j=\rank\bm X\equiv m.
 \label{eq:sumrule}
\end{equation}
The ports are fixed before the task is chosen, after which the coherent probe
is reoptimized.  Equation~\eqref{eq:sumrule} turns the projector trace into a
Fisher-retention budget: at fixed reachable rank, non-Hermitian preparation
redistributes that budget across a complete task basis without changing its
total.  Partial task sets
obey the projected trace identity given in Sec.~S2C of the Supplemental
Material.

\begin{figure}[!t]
 \centering
 \includegraphics[width=0.98\columnwidth]{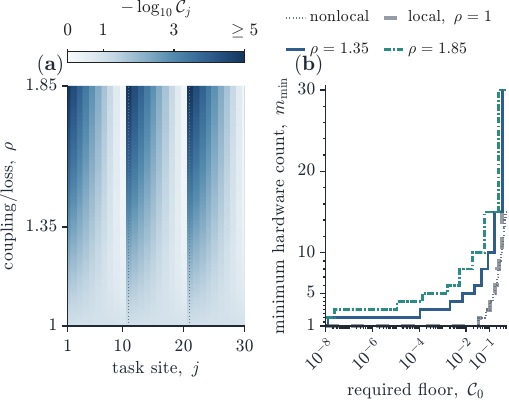}
\caption{\textbf{Local-port law for Fisher retention.}
 For an on-resonance feed-forward chain with \(\bm Q=\bm I\) and \(N=30\):
 (a) Task-specific retention loss \(-\log_{10}\Ccal_j\), clipped at five
 decades, versus \(\rho=\kappa/\gamma\) for local ports
 \(\{1,11,21\}\).  Thin lines mark the internal port boundaries.  Each slice
 obeys \(\sum_j\Ccal_j=3\), and each segment sums to unity; increasing \(\rho\)
 shifts that unit toward the downstream end.
 (b) Minimum nonlocal rank or local-port count required to guarantee
 \(\Ccal_j\ge\Ccal_0\).  The nonlocal curve is the attainable Schur--Horn
 reference \(\lceil N\Ccal_0\rceil\); the local curves follow
 Eq.~\eqref{eq:localBound} and its neutral-buildup limit.  Each upward step
 corresponds to one additional physical port.}
 \label{fig:jordan}
\end{figure}

To obtain a closed-form interval law, we now specialize to uniform
nearest-neighbor forward coupling, uniform loss, on-resonance operation,
\(\bm Q=\bm I\), onsite rank-one tasks, and fixed downstream measurement maps.
Taking \(\widetilde{\bm w}_j=\bm e_j\), let the local ports be
\(\bm B_{\mathcal P}=(\bm e_{p_1},\ldots,\bm e_{p_m})\), with
\(1=p_1<\cdots<p_m\), and set \(p_{m+1}=N+1\) and
\(L_\ell=p_{\ell+1}-p_\ell\).  The associated site interval is
\(I_\ell=\{p_\ell,\ldots,p_{\ell+1}-1\}\).  At \(\omega=0\), the Green column launched at
site \(p\) is
\begin{equation}
 \bm g_p=\bm G\bm e_p
 =\frac{\ii}{\gamma}\sum_{r=0}^{N-p}(-\ii\rho)^r\bm e_{p+r}.
 \label{eq:GreenColumnMain}
\end{equation}
With \(\bm g_{N+1}=0\), eliminating the next port column gives
\begin{equation}
 \bm g_{p_\ell}-(-\ii\rho)^{L_\ell}\bm g_{p_{\ell+1}}
 =\frac{\ii}{\gamma}\sum_{r=0}^{L_\ell-1}
 (-\ii\rho)^r\bm e_{p_\ell+r}.
 \label{eq:segmentEliminationMain}
\end{equation}
These vectors have disjoint supports and span the original Green columns.
Normalizing them therefore gives, for \(p_\ell\le j<p_{\ell+1}\),
\begin{equation}
 \Ccal_j=
 \frac{\rho^{2(j-p_\ell)}}{\sum_{r=0}^{L_\ell-1}\rho^{2r}},
 \qquad \rho>0.
 \label{eq:segment}
\end{equation}
Up to a diagonal propagation-phase gauge, the corresponding orthonormal
segment vector has components proportional to \(\rho^{j-p_\ell}\) on its
own segment and vanishes elsewhere.  Hence the sitewise retentions in each
nonempty segment sum to unity.  Balancing segment lengths gives the
local-port max--min result
\begin{equation}
 \Ccal_{\min}^{\rm loc,\star}
 =\frac{\rho^2-1}
 {\rho^{2\lceil N/m\rceil}-1}
 \qquad(\rho>1).
 \label{eq:localBound}
\end{equation}
Figure~\ref{fig:jordan} summarizes the spatial redistribution of the fixed-rank
budget and the local-port count required to maintain a retention floor.

The exponential factor is the cost of assigning an interval of length
\(\lceil N/m\rceil\) to one local port: directional propagation concentrates
the interval's conserved retention unit at its downstream end.
At the neutral-buildup point \(\rho=1\), the chain remains strictly
one-way.  Equation~\eqref{eq:segment} then gives
\(\Ccal_j=1/L_\ell\) and
\(\Ccal_{\min}^{\rm loc,\star}=1/\lceil N/m\rceil\).  If the rank-\(m\)
internal-control subspace can be designed without locality constraints, the
uniform Schur--Horn point \(\Ccal_j=m/N\) is attainable
~\cite{Schur1923,Horn1954,HornJohnson2013}.  For
\(\rho>1\), guaranteeing \(\Ccal_j\ge\Ccal_0\), with
\(0<\Ccal_0\le1\), requires
\begin{equation}
 L_\ell\le L_{\max}=\left\lfloor
 \frac{\ln[1+(\rho^2-1)/\Ccal_0]}{2\ln\rho}\right\rfloor,
 \qquad m\ge\left\lceil\frac{N}{L_{\max}}\right\rceil.
 \label{eq:portDensity}
\end{equation}
Section~S4 of the Supplemental Material treats \(\rho=0\), diagonal resource
metrics, and localized response columns, and relates the monochromatic Fisher
ratio to finite-time control-energy bounds
~\cite{Chen2016EnergyScaling,Wang2017PhysicalControllability,
ZhaoPasqualetti2018,KlicksteinSorrentino2020Gramian,
Alizadeh2023LongestChain,Nazerian2026Frequency,Duan2022LocalizedControl}.

\begin{figure}[!t]
 \centering
 \includegraphics[width=0.98\columnwidth]{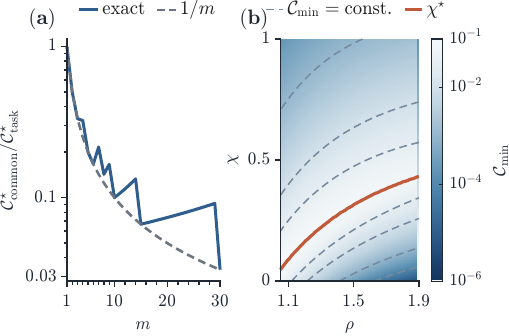}
 \caption{\textbf{Probe order and recovery of spatial coverage beyond the
 strictly feed-forward limit.}
 (a) Fraction of the task-specific worst-site retention preserved by one common
 probe for a balanced \(N=30\) chain at \(\rho=1.35\).  The dashed line is
 \(1/m\), attained when the \(m\) port segments have equal length; deviations
 reflect the integer segment imbalance in Eq.~\eqref{eq:commonProbeExact}.
 (b) Worst-site retention \(\Ccal_{\min}(\rho,\chi)\) for \(N=30\) and ports
 \(\{1,11,21\}\).  Dashed contours mark fixed retention levels and the solid
 curve follows the maximizing \(\chi^\star(\rho)\); \(\chi=0\) and \(1\)
 denote the feed-forward and reciprocal limits.  All plotted points are
 stable.}
 \label{fig:protocolRobustness}
\end{figure}

The order of target selection and probe optimization changes the worst-case
value.  For unit internal resource, define the task-specific and common-probe
retentions by
\begin{align}
 \Ccal_{\rm task}^{\star}
 &=\min_j\max_{\substack{\bm z\in\ran\bm X\\ \norm{\bm z}\le1}}
 |\bm e_j^\dagger\bm z|^2,\nonumber\\
 \Ccal_{\rm common}^{\star}
 &=\max_{\substack{\bm z\in\ran\bm X\\ \norm{\bm z}\le1}}
 \min_j|\bm e_j^\dagger\bm z|^2.
 \label{eq:taskCommonRetention}
\end{align}
Let \(d_\ell=\min_{p_\ell\le j<p_{\ell+1}}\Ccal_j\).  The disjoint segment
basis gives the exact pair
\begin{equation}
 \Ccal_{\rm task}^{\star}=\min_\ell d_\ell,
 \qquad
 \Ccal_{\rm common}^{\star}=\left(\sum_{\ell=1}^{m}d_\ell^{-1}\right)^{-1}.
 \label{eq:commonProbeExact}
\end{equation}
For equal segment lengths, the common-probe optimum incurs the additional factor
\(\Ccal_{\rm common}^{\star}=\Ccal_{\rm task}^{\star}/m\).
Figure~\ref{fig:protocolRobustness}(a) displays this common-probe penalty and
the small deviations caused by integer segment imbalance.

To examine how coverage changes beyond the strictly feed-forward limit, we add
reverse hopping,
\begin{equation}
 \bm H_\chi=-\ii\gamma\bm I+\kappa\bm T_+
             +\chi\kappa\bm T_-,\qquad \bm T_-=\bm T_+^{\mathsf T}.
 \label{eq:reverseHopping}
\end{equation}
Here \(\chi\) is dimensionless: \(\chi=0\) is the one-way chain and
\(\chi=1\) is the reciprocal limit.  Figure~\ref{fig:protocolRobustness}(b)
maps the worst-site retention over \(1.05\le\rho\le1.9\) and
\(0\le\chi\le1\) for \(N=30\), \(\omega=0\), \(\gamma=1\), and ports
\(\{1,11,21\}\).  The hopping matrix has a real spectrum, so the entire
plotted domain retains the decay rate \(\gamma\).  A finite reverse coupling
can recover much of the
coverage lost at large \(\rho\), but the maximizing \(\chi\) is generally
neither zero nor one.  The map shows how backward propagation
reshapes the reachable subspace between the one-way and reciprocal limits.
Section~S2 of the Supplemental Material gives the associated local
rank-preserving perturbation bound.

\section{Source--internal-resource crossover}

The source and internal comparisons are the two endpoints of a joint-resource
problem.  For a scalar target, let \(\bm R\succeq0\) be the source metric and define
\begin{equation}
 \bm H=\bm X^\dagger\bm X,\qquad
 \bm K=\bm X^\dagger\bm A^\dagger\bm A\bm X.
 \label{eq:KH}
\end{equation}
The optimization
\begin{equation}
 \max_{\bm s}\;2\bm s^\dagger\bm K\bm s,\quad
 \bm s^\dagger\bm R\bm s\le P_0,\quad
 \bm s^\dagger\bm H\bm s\le U_0
 \label{eq:jointPrimal}
\end{equation}
is finite precisely when
\begin{equation}
 \ker\bm R\cap\ker\bm H\subseteq\ker\bm K.
 \label{eq:jointFinite}
\end{equation}
For positive budgets its dual is~\cite{BeckEldar2006}
\begin{align}
 F_{\mu,P,U}^{\star}
 &=2\min_{\lambda_P,\lambda_U\ge0}
 (\lambda_P P_0+\lambda_U U_0),
 \label{eq:jointDualA}\\
 \bm K&\preceq\lambda_P\bm R+\lambda_U\bm H.
 \label{eq:jointDualB}
\end{align}
At differentiable points,
\begin{equation}
 \partial_{P_0}F_{\mu,P,U}^{\star}=2\lambda_P^\star,\qquad
 \partial_{U_0}F_{\mu,P,U}^{\star}=2\lambda_U^\star,
 \label{eq:shadow}
\end{equation}
so the dual variables are the marginal Fisher values of the two resources.

When \(\bm R\) is positive on the admitted input space, introduce the source-whitened preparation Gramian and its resource filter,
\begin{equation}
 \bm W_0=\bm X\bm R^{-1}\bm X^\dagger,\qquad
 \bm W_\tau=\bm W_0(\bm I+\tau\bm W_0)^{-1}.
 \label{eq:resourceFilter}
\end{equation}
The joint optimum is the one-parameter envelope
\begin{equation}
 F_{\mu,P,U}^{\star}=2\inf_{\tau\ge0}
 (P_0+\tau U_0)\lambda_{\max}
 (\bm A\bm W_\tau\bm A^\dagger).
 \label{eq:spectralEnvelope}
\end{equation}
Each preparation eigenvalue \(w_i\) is transformed to \(w_i/(1+\tau w_i)\).
The source-limited endpoint retains the preparation spectrum.  In the opposite limit,
\begin{equation}
 \lim_{\tau\rightarrow\infty}\tau\bm W_\tau=\bm\Pi_X
 \label{eq:filterLimit}
\end{equation}
recovers the fixed-\(U\) projector.  The frontier is source limited
when \(\lambda_U^\star=0\), mixed when both multipliers are positive, and
 internal-resource limited when \(\lambda_P^\star=0\).  For semidefinite
 \(\bm R\), zero-cost directions require the support conditions derived in
 Sec.~S3 of the Supplemental Material, which also gives rank-one recovery and
 the transition conditions.

For the one-port feed-forward chain with \(P_0,U_0>0\), the preparation
buildup is
\begin{equation}
 \beta_N=\gamma^{-2}\sum_{r=0}^{N-1}\rho^{2r},
 \label{eq:betaN}
\end{equation}
and the Fisher ratio with both constraints, measured relative to the
full-state drive benchmark at fixed $U$, becomes
\begin{equation}
\zeta_j(P_0,U_0)\equiv
\frac{F_{\mu,P,U,B,j}^{\star}}
{F_{\mu,U,\rm full,j}^{\star}}
=\Ccal_j\min\left[1,\frac{P_0}{U_0}\beta_N\right].
 \label{eq:jointChain}
\end{equation}
The internal-resource-limited branch saturates at \(\Ccal_j\).  The
source-limited branch retains the preparation buildup.  Increasing \(\rho>1\)
therefore lowers the emission-normalized preparation cost of a selected
downstream task but reduces the worst-site retention between sparse local ports.
For one effective preparation direction, Eq.~\eqref{eq:jointChain} joins these
two branches directly.  Multiple drive directions generate an intervening
mixed regime in which both budgets carry positive marginal Fisher value.

Section~S3 of the Supplemental Material resolves the
internal-resource-limited, mixed, and source-limited branches in a stable
four-mode system.  The pole-identical dimers in Sec.~S6B and Fig.~S6 show that
identical complex pole trajectories can nevertheless produce different
fixed-resource Fisher frontiers.

\section{Reduced-order noisy RLC/VCCS network}

\begin{figure}[t]
 \centering
 \includegraphics[width=0.98\columnwidth]{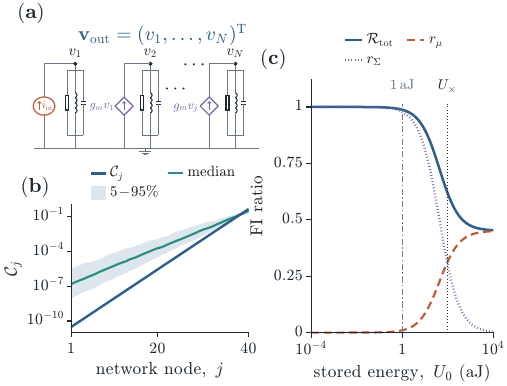}
 \caption{\textbf{Reduced-order noisy RLC/VCCS network.}
 (a) Each VCCS is connected between the downstream node and the reference node
 and injects \(g_m v_j\), controlled by the upstream voltage \(v_j\).  The
 coherent source \(i_{\rm in}\) drives node 1, and the monitored voltages form
 \(\bm v_{\rm out}=\bm Z\bm B\bm s\).
 (b) Mean-FI retention of local-port driving relative to ideal full-state
 driving at fixed coherent mean stored energy.  Curves show the nominal network
 and disorder median; shading is the
 5--95\% interval over 300 realizations with 5\% component disorder.
 (c) Edge-task total retention
 \(\mathcal R_{\rm tot}=r_\mu+r_\Sigma\), with
 \(r_\mu=F_{\mu,U,B}^{\star}/(F_{\mu,U,\rm full}^{\star}+F_\Sigma)\) and
 \(r_\Sigma=F_\Sigma/(F_{\mu,U,\rm full}^{\star}+F_\Sigma)\).
 The guides mark \(U_0=1~\mathrm{aJ}\) and
 \(U_\times=\RLCMeanCovarianceCrossoverAJ~\mathrm{aJ}\), where the physical-port
 mean and covariance contributions are equal.  The common covariance term
 controls the low-\(U_0\) limit: \(r_\Sigma\to1\) while \(r_\mu\to0\), so
 \(\mathcal R_{\rm tot}\to1\) is a covariance-only baseline rather than high
 mean-response retention.}
 \label{fig:network}
\end{figure}

Non-Hermitian spectral response has been demonstrated in topolectrical and
integrated circuits~\cite{Yuan2023Circuit,Deng2024Circuit}.  An exceptional
point has also been realized in two LC resonators linked by a unidirectional
voltage-follower coupler~\cite{Zhao2024}.
Our unilateral RLC/VCCS chain gives a reduced-order circuit-level
illustration of the finite-port comparison using a coherent mean
electromagnetic-energy metric.
Figure~\ref{fig:network}(a) shows the physical model.  A coherent current drives
node 1, and each node is a parallel RLC resonator.  A VCCS connected between
node \(j+1\) and the reference node injects the current \(g_m v_j\), controlled
by the upstream voltage \(v_j\).  All node voltages are monitored.

The reduced model uses high-impedance voltage readout at all 40 nodes.  For
\(N=40\) parallel RLC nodes, the admittance is
\begin{equation}
 \bm Y(\Omega)=
 \diag\!\left[
 R^{-1}+\ii\Omega C_j+(\ii\Omega L_j)^{-1}
 \right]
 -\sum_{j=1}^{N-1}g_j|j+1\rangle\langle j|.
 \label{eq:Y}
\end{equation}
For the coherent mean voltage phasor \(\bar{\bm v}=\bm Y^{-1}\bm B\bm s\),
the constrained resource is
\(U_{\rm coh}=\bar{\bm v}^{\dagger}\bm Q\bar{\bm v}\), with
\(\bm Q=\bm C/4+\bm L^{-1}/(4\Omega^2)\).  Thus \(U_{\rm coh}\) measures the
coherent mean electromagnetic energy.  The calibrated likelihood contains
300-K resistor Johnson--Nyquist noise, a 600-K phenomenological effective
active-source bath, and colored correlated readout noise in a 1-Hz
noise-equivalent bandwidth~\cite{Johnson1928,Nyquist1928}.  The metric is fixed
at its nominal component values when evaluating the local Fisher derivative.
The nominal values are
\begin{equation}
 C=3.2~\mathrm{nF},\quad
 L=20~\mathrm{nH},\quad
 R=20~\Omega,\quad
 g_m=0.0675~\mathrm{S},
 \label{eq:circuitvalues}
\end{equation}
giving $f_0=19.894$ MHz and $g_m R=1.35$.  The feed-forward admittance is
triangular, and its stability margin follows from the roots of each local
parallel-RLC polynomial.  Both the nominal network and every sampled
realization are underdamped and stable within this reduced model.

For a local capacitance task, the constrained mean-response retention
\begin{equation}
 \Rcal_{\mu,j}\equiv
 \frac{F_{\mu,U,B,j}^{\star}}{F_{\mu,U,\rm full,j}^{\star}}
 =\Ccal_j
 \label{eq:networkRatio}
\end{equation}
follows the reachable-projector law.  Figure~\ref{fig:network}(b) shows its
spatial form and its response to component disorder.  For 300 realizations
with \(5\%\) independent component disorder, the
skin-edge mean-FI retention has median \(\RLCMedian\) and a 5--95\% interval
from \(\RLCQFive\) to \(\RLCQNinetyFive\).

The capacitance perturbation also changes the output covariance, adding
\(F_\Sigma\) independently of the coherent drive.  Figure~\ref{fig:network}(c)
resolves the crossover.  Writing the full-state-drive mean block as
\(F_{\mu,U,\mathrm{full}}^\star=a_N U_0\), the edge-task total retention is
\begin{equation}
 \Rcal_{\rm tot}(U_0)=
 \frac{\Ccal_N a_N U_0+F_\Sigma}{a_N U_0+F_\Sigma}.
 \label{eq:networkTotalRetention}
\end{equation}
At \(U_0=1~\mathrm{aJ}\), the skin-edge mean-FI
retention is \(\RLCNominalEdge\), while the total retention is
\(\RLCNominalTotal\).  The physical-port mean block equals the covariance block
at \(U_\times=\RLCMeanCovarianceCrossoverAJ~\mathrm{aJ}\).  Below this scale,
the shared covariance information drives the total ratio toward unity; above
it, the coherent mean response takes over and the ratio approaches
\(\Ccal_N\).  Sections~S6 and S7 of the Supplemental Material give the full
noise construction and numerical conditioning analysis.

\section{Discussion and conclusion}

When drive hardware precedes task selection, finite local ports impose a
spatial information cost distinct from non-normal amplification.  At fixed
internal resource, the preparation
singular values disappear from the optimized mean-FI region and only the
reachable subspace remains.  In a feed-forward chain this geometry becomes an
interval conservation law: each local port supplies one unit of taskwise
retention to the sites in its interval.  Directionality concentrates that unit
downstream, improving a selected task while exponentially reducing the worst
task when one port serves many sites.

The resulting design variable is the largest interval assigned to a port, or
equivalently the port density required to maintain a retention floor.  A probe
shared across all task locations carries an additional, exactly calculable
penalty.  When source power is constrained as well, the preparation Gramian
returns through the joint-resource crossover and quantifies the cost of
reaching the retained fields.

The RLC/VCCS model carries this distinction to stored electromagnetic energy
and shows how covariance information enters alongside the coherent mean
response.  The pole-identical dimer comparison further shows that spectral
sensitivity and operational Fisher information need not follow the same
ordering.  Once the input--output and resource matrices are calibrated, the
same projector construction applies to integrated photonic, microwave,
acoustic, mechanical, and other linear wave networks with finite local access.
The feed-forward result provides an analytically closed benchmark for these
systems: directional preparation can save source power for a chosen target,
while spatially robust sensing requires enough ports to cover tasks selected
after hardware installation.

\begin{acknowledgments}
OpenAI Codex (GPT-5) assisted with debugging selected MATLAB routines,
checking numerical diagnostics, and language editing.

This work was supported by the Leading-Edge Technology Program of Jiangsu
Natural Science Foundation (No.~BK20232001); the Fundamental Research Funds for
the Central Universities (Nos.~0213-14380292 and 0210-14380286); the National
Natural Science Fund for Excellent Young Scholars; and the Nanjing University
Integrated Research Platform of the Ministry of Education--Top Talents Program.
\end{acknowledgments}

\section*{Data Availability}

The numerical data and MATLAB code supporting this study are not publicly
available during peer review because the associated repository is under
preparation.  They are available from the contact author upon reasonable
request and will be deposited in a persistent repository upon acceptance.  No
experimental data were generated.

\nocite{ArandesBergholtz2025,BjorckGolub1973,Blackwell1953,
 BraunsteinCaves1994,Caves1982,Clerk2010,Ding2023,Drineas2012,
 HatanoNelson1996,KiorpelidisMakris2025,KochBudich2022,Kunst2018,
 Mahoney2011,Okuma2020,Sarkar2024Critical,SuSchriefferHeeger1979,
 Wang2021CPA,Wang2026CPAEP3,Wedin1972,Wiersig2020,Xiao2019,
 YaoWang2018,YokomizoMurakami2019,Zamir1998,Zhang2019}

\def\bibinfo#1#2{%
  \ifstrequal{#1}{title}{\textcolor{black}{#2}}{#2}}
\hypersetup{urlcolor=blue}
\bibliography{refs}
\end{document}


\title{Supplemental Material for ``Fisher-information retention under local driving in non-Hermitian feed-forward chains''}
\author{Qingrui Bai}
\affiliation{School of Electronic Science and Engineering, Nanjing University, Nanjing 210023, China}
\affiliation{School of Integrated Circuits, Nanjing University, Suzhou 215163, China}
\affiliation{National Key Laboratory of Transient Impact, Nanjing 210023, China}
\author{Zhichao Li}
\affiliation{School of Electronic Science and Engineering, Nanjing University, Nanjing 210023, China}
\affiliation{School of Integrated Circuits, Nanjing University, Suzhou 215163, China}
\affiliation{National Key Laboratory of Transient Impact, Nanjing 210023, China}
\author{Junlong Kou}
\email[Contact author: ]{jlkou@nju.edu.cn}
\affiliation{School of Electronic Science and Engineering, Nanjing University, Nanjing 210023, China}
\affiliation{School of Integrated Circuits, Nanjing University, Suzhou 215163, China}
\affiliation{National Key Laboratory of Transient Impact, Nanjing 210023, China}
\affiliation{Key Laboratory of Intelligent Optical Sensing and Manipulation, Ministry of Education, Nanjing University, Nanjing 210093, China}
\affiliation{Jiangsu Key Laboratory of Semiconductor Laser and Sensing Technology, Suzhou 215163, China}
\affiliation{Wujin-NJU Institute of Future Technology, Changzhou 213153, China}
\maketitle
\onecolumngrid

\setcounter{equation}{0}\renewcommand{\theequation}{S\arabic{equation}}
\setcounter{figure}{0}\renewcommand{\thefigure}{S\arabic{figure}}
\setcounter{table}{0}\renewcommand{\thetable}{S\arabic{table}}
\setcounter{section}{0}\renewcommand{\thesection}{S\arabic{section}}

\begingroup
\small
\setcounter{tocdepth}{1}
\makeatletter
\let\l@subsubsection\@gobbletwo
\makeatother
\tableofcontents
\endgroup
\clearpage

\suppsection{Conventions and statistical model}
\label{sec:scope}

\suppsubsection{Notation, stability, and input--output convention}
\label{sec:notation}

The internal state is $\bm a\in\mathbb C^N$, the controlled input is
$\bm s\in\mathbb C^p$, and the monitored output is
$\bm b\in\mathbb C^q$.  At a fixed real frequency,
\begin{equation}
 \bm b=\bm S(\bm\theta,\omega)\bm s+\bm n,
 \qquad
 \bm S=\bm D-\ii\bm C\bm G\bm B,
 \qquad
 \bm G=(\bm H_{\rm eff}-\omega\bm I)^{-1}.
 \label{eq:ioSupp}
\end{equation}
The operating point is stable and $\bm G$ exists.  Under the Fourier convention
used in Eq.~\eqref{eq:ioSupp}, all poles of the causal dynamics lie in the
 stable half-plane.  A real-frequency resolvent may still be non-normal
and ill conditioned.

We first take a real parameter $\epsilon$ encoded through the internal generator,
\begin{equation}
 \bm V\equiv\partial_\epsilon\bm H_{\rm eff},
 \qquad
 \partial_\epsilon\bm G=-\bm G\bm V\bm G,
 \qquad
 \partial_\epsilon\bm S=\ii\bm C\bm G\bm V\bm G\bm B.
 \label{eq:dSsupp}
\end{equation}
Parameter dependence of $\bm B$, $\bm C$, $\bm D$, or a noise coupling adds
the corresponding product-rule terms.  The external-resource theorem uses the
resulting $\partial_\epsilon\bm S$.  The fixed-\(U\) factorization
below applies to the generator-encoded tangent in Eq.~\eqref{eq:dSsupp}.

The notation used in the derivations is summarized below.
\begin{center}
\small
\begin{tabular}{@{}p{3.6cm}p{10.6cm}@{}}
\toprule
Symbol & Meaning \\
\midrule
$\bm G$ & Stable real-frequency resolvent \\
$\bm B,\bm C,\bm D$ & Controlled input coupling, monitored output coupling, and direct path \\
$\bm V$ & Parameter tangent $\partial_\epsilon\bm H_{\rm eff}$ \\
$\bm\Sigma_c$ & Proper complex output covariance for one statistically independent record \\
$\bm s,\bm a,\bm b$ & Peak-amplitude complex phasors for input, internal state, and monitored output; an RMS convention rescales the quadratic budgets consistently \\
$\bm S_n^{(1)}$ & One-sided output noise spectral-density matrix \\
$\bm R\succeq0$ & External source-resource metric \\
$\bm Q\succ0$ & Calibrated internal quadratic-resource metric \\
$P_0$ & Source-resource budget, with units set by $\bm R$ (power for power-wave input amplitudes) \\
$U_0$ & Internal-resource budget, with units set by $\bm Q$ (joules for the RLC stored-energy metric) \\
$\bm X=\bm Q^{1/2}\bm G\bm B$ & Map from drive amplitudes to the resource-normalized internal field \\
$\bm\Pi_X=\bm X\bm X^+$ & Orthogonal projector onto $\ran\bm X$ \\
$\Ccal_j$ & Fraction of the ideal full-state-drive mean FI retained for task $j$ \\
$\Ecal_j$ & Emission to the monitored outputs after noise whitening \\
$\beta$ & Buildup from the source to the internal field \\
\bottomrule
\end{tabular}
\end{center}

Throughout the general theory, \(U\) denotes a calibrated internal quadratic resource,
\begin{equation}
 U=\bm a^\dagger\bm Q\bm a,
 \qquad \bm Q\succ0.
 \label{eq:energyLedger}
\end{equation}
It may represent modal occupation, a saturation-weighted cost, or another
calibrated internal quantity.  In the RLC example it is the time-averaged stored
electromagnetic energy of the coherent mean phasor.  Pump and readout power
belong to $\bm R$ or to additional constraints.  For a record of equivalent
noise bandwidth $B_{\rm eq}$,
\(\bm\Sigma_c=\bm S_n^{(1)}B_{\rm eq}\); Fisher matrices add over independent
records.  The RLC calculation uses $B_{\rm eq}=1$ Hz and one calibrated complex
record.

For the closed-form feed-forward chain, the local drive ports and local tasks
share the site basis at a stable real frequency.  The port locations are fixed,
and the complex drive amplitudes are optimized after the task site is chosen.
The analytic expressions use \(\bm Q=\bm I\); a positive diagonal metric
 reweights the corresponding segment profiles.  The ideal full-state drive
 benchmark replaces \(\bm B\) with \(\bm I\) while keeping the likelihood and
 internal budget fixed.

\suppsubsection{Proper complex Gaussian Fisher information}
\label{sec:gaussianFI}

For a proper complex Gaussian observation $\bm z\sim\mathcal{CN}(\bm\mu,\bm\Sigma_c)$,
\begin{equation}
 p(\bm z|\epsilon)=\frac{\exp[-(\bm z-\bm\mu)^\dagger\bm\Sigma_c^{-1}(\bm z-\bm\mu)]}
 {\pi^q\det\bm\Sigma_c}.
 \label{eq:complexpdf}
\end{equation}
Differentiating the log likelihood with respect to a real parameter gives
\begin{align}
 \partial_\epsilon\ell={}&2\operatorname{Re}\!\left[(\partial_\epsilon\bm\mu)^\dagger
 \bm\Sigma_c^{-1}(\bm z-\bm\mu)\right] \nonumber\\
 &+(\bm z-\bm\mu)^\dagger\bm\Sigma_c^{-1}(\partial_\epsilon\bm\Sigma_c)
 \bm\Sigma_c^{-1}(\bm z-\bm\mu)
 -\Tr(\bm\Sigma_c^{-1}\partial_\epsilon\bm\Sigma_c).
 \label{eq:score}
\end{align}
Using the proper complex Gaussian moments and standard Fisher-information identities~\cite{Kay1993,VanTrees2001,SchreierScharf2010}, the mean and covariance scores are uncorrelated, and
\begin{equation}
 F_{\epsilon\epsilon}=2(\partial_\epsilon\bm\mu)^\dagger\bm\Sigma_c^{-1}
 (\partial_\epsilon\bm\mu)
 +\Tr[(\bm\Sigma_c^{-1}\partial_\epsilon\bm\Sigma_c)^2].
 \label{eq:complexFI}
\end{equation}
For several real parameters $\theta_\alpha$,
\begin{align}
 F_{\alpha\beta}={}&2\operatorname{Re}\!\left[(\partial_\alpha\bm\mu)^\dagger\bm\Sigma_c^{-1}
 (\partial_\beta\bm\mu)\right]\nonumber\\
 &+\Tr[\bm\Sigma_c^{-1}(\partial_\alpha\bm\Sigma_c)
 \bm\Sigma_c^{-1}(\partial_\beta\bm\Sigma_c)].
 \label{eq:Fmatrix}
\end{align}
With $\bm\mu=\bm S\bm s$ and probe-independent covariance, the
coherent-input-dependent block is
\begin{equation}
 F_\mu(\bm s)=2\bm s^\dagger(\partial_\epsilon\bm S)^\dagger
 \bm\Sigma_c^{-1}(\partial_\epsilon\bm S)\bm s.
 \label{eq:Fmusupp}
\end{equation}
We refer to the first term as mean FI.  The covariance FI
\begin{equation}
 F_\Sigma=\Tr[(\bm\Sigma_c^{-1}\partial_\epsilon\bm\Sigma_c)^2]
 \label{eq:Fsigma}
\end{equation}
is additive and independent of the coherent amplitude when the noise itself is
probe independent.  The port-rank retention budget constrains the mean FI.
Section~\ref{sec:covNuisance} derives the covariance contribution and the
resulting crossover.

For $M$ identically prepared independent records, the total Fisher matrix is
$M$ times the per-record Fisher matrix.  A multifrequency experiment with
independent bins has $F_{\rm total}=\sum_\nu M_\nu F(\omega_\nu)$.  Comparing a
scan with a single-frequency measurement uses equal total sample number or
integration time.

\suppsubsubsection{Efficient-score geometry}

For each real parameter define the whitened mean and covariance tangents
\begin{equation}
 \bm u_\alpha=\bm\Sigma_c^{-1/2}\partial_\alpha\bm\mu,
 \qquad
 \bm Z_\alpha=\bm\Sigma_c^{-1/2}(\partial_\alpha\bm\Sigma_c)
 \bm\Sigma_c^{-1/2}.
 \label{eq:scoreComponentsSupp}
\end{equation}
 The covariance tangent is Hermitian.  On the real Hilbert-space direct sum of
 complex vectors and Hermitian matrices, introduce
\begin{equation}
 \left\langle(\bm u,\bm U),(\bm v,\bm V)\right\rangle_{\rm sc}
 =2\operatorname{Re}(\bm u^\dagger\bm v)+\Tr(\bm U\bm V).
 \label{eq:scoreInnerProduct}
\end{equation}
With \(\bm g_\alpha=(\bm u_\alpha,\bm Z_\alpha)\), Eq.~\eqref{eq:Fmatrix} becomes
\begin{equation}
 F_{\alpha\beta}=\langle\bm g_\alpha,\bm g_\beta\rangle_{\rm sc}.
 \label{eq:FisherScoreGram}
\end{equation}
The proper complex Gaussian Fisher matrix is a score Gram matrix.  Its mean and
covariance sectors are orthogonal before nuisance elimination.

Let \(t\) denote one target parameter and let \(n\) collect the nuisance parameters.  For a positive-semidefinite nuisance-prior precision \(\bm P_n\), define
\begin{align}
 \Delta_{\bm c}\bm\mu
 &=\partial_t\bm\mu-\sum_{j\in n}c_j\partial_j\bm\mu,\nonumber\\
 \Delta_{\bm c}\bm\Sigma_c
 &=\partial_t\bm\Sigma_c-\sum_{j\in n}c_j\partial_j\bm\Sigma_c.
 \label{eq:scoreResiduals}
\end{align}
Completion of the square in the Gram matrix gives
\begin{align}
 F_{t|n}=\min_{\bm c\in\mathbb R^{|n|}}\bigl\{&
 2\norm{\bm\Sigma_c^{-1/2}\Delta_{\bm c}\bm\mu}^2
 +\norm{\bm\Sigma_c^{-1/2}\Delta_{\bm c}\bm\Sigma_c
 \bm\Sigma_c^{-1/2}}_F^2\nonumber\\
 &+\bm c^{\mathsf T}\bm P_n\bm c\bigr\}.
 \label{eq:efficientScoreDistanceSupp}
\end{align}
A minimum-norm solution is
\(\bm c_\star=(\bm F_{nn}+\bm P_n)^+\bm F_{nt}\) under the support condition.
Equation~\eqref{eq:efficientScoreDistanceSupp} is the generalized Schur
complement: the augmented target score is projected away from the nuisance
score span.  The same coefficients act in the mean and covariance sectors, so
their Fisher blocks are summed before this projection.

For a feasible probe set \(\mathcal S\), the nuisance-adjusted scalar
optimum is
\begin{align}
 F_{t|n}^{\star}=\max_{\bm s\in\mathcal S}\min_{\bm c\in\mathbb R^{|n|}}\bigl\{&
 2\norm{\bm\Sigma_c^{-1/2}\Delta_{\bm c}\bm\mu(\bm s)}^2
 +\norm{\bm\Sigma_c^{-1/2}\Delta_{\bm c}\bm\Sigma_c
 \bm\Sigma_c^{-1/2}}_F^2\nonumber\\
 &+\bm c^{\mathsf T}\bm P_n\bm c\bigr\}.
 \label{eq:operationalMaxMinSupp}
\end{align}
Equation~\eqref{eq:operationalMaxMinSupp} keeps nuisance elimination inside
the probe design.  A parameter-independent Markov readout maps each score to a
conditional expectation and therefore contracts both the Fisher matrix and
the efficient-score distance~\cite{Blackwell1953,Zamir1998}; invertible
parameter-independent coordinate changes preserve them.  Quadratic resources
act on the probe-dependent mean score.  A probe-independent covariance score
retains its per-record value.

\suppsubsection{Measurement design and coherent probes}
\label{sec:completeProtocol}
\label{sec:ensembleDesign}

\emph{Comparison protocol.} A comparison fixes the monitored outputs,
covariance, allowed probes, resources, record count, and nuisance parameters.
When \(\bm\Sigma_c\) is probe
independent, the scalar mean-response information for the likelihood in
Eq.~\eqref{eq:complexpdf} is
\begin{equation}
 F_\mu(\bm s)=2\bm s^\dagger\bm M_\theta\bm s,
 \qquad
 \bm M_\theta=(\partial_\theta\bm S)^\dagger
 \bm\Sigma_c^{-1}(\partial_\theta\bm S).
 \label{eq:FisherOperatorSupp}
\end{equation}
For two architectures on the same finite-cost support, architecture A dominates
the mean-response block for every admissible probe precisely when
\begin{equation}
 \bm R^{+1/2}(\bm M_{\theta,A}-\bm M_{\theta,B})
 \bm R^{+1/2}\succeq0.
 \label{eq:dominanceSupp}
\end{equation}
If this difference is indefinite, the two architectures exchange order across
admissible probes.  Changing the monitored outputs or applying a lossy
post-processing map changes the statistical experiment.  The covariance-derivative
term enters before nuisance elimination and can alter the total-Fisher ordering.

A collection of coherent probes \(\{\bm s_k\}\), repeated \(n_k\) times, is represented by
\begin{equation}
 \bm\Xi=\sum_k n_k\bm s_k\bm s_k^\dagger\succeq0.
 \label{eq:XiSupp}
\end{equation}
Writing \(\bm T_\alpha=\bm\Sigma_c^{-1/2}
\partial_{\vartheta_\alpha}\bm S\), its mean-FI matrix is
\begin{equation}
 [\bm F_\mu(\bm\Xi)]_{\alpha\beta}
 =
 2\operatorname{Re}\operatorname{Tr}
 (\bm\Xi\,\bm T_\alpha^\dagger\bm T_\beta).
 \label{eq:ensembleFIMSupp}
\end{equation}
If one record contributes the covariance block \(\bm F_\Sigma^{(1)}\), then
\begin{equation}
 \bm F_{\rm data}(\bm\Xi)
 =\bm F_\mu(\bm\Xi)+N_{\rm rec}\bm F_\Sigma^{(1)},
 \qquad N_{\rm rec}=\sum_k n_k.
 \label{eq:ensembleFullSupp}
\end{equation}
Quadratic source and internal resources become affine constraints,
\begin{equation}
 \operatorname{Tr}(\bm R\bm\Xi)\le P_{\rm tot},
 \qquad
 \operatorname{Tr}(\bm H\bm\Xi)\le U_{\rm tot}.
 \label{eq:ensembleResourcesSupp}
\end{equation}
Three designs recur below.  A task-specific design reoptimizes \(\bm s\) after
the target direction is specified.  A common-probe design uses one \(\bm s\)
for every candidate target.  A probe ensemble optimizes \(\bm\Xi\) and allocates
records among several coherent settings.  The local retention law uses the
first design; unknown-location sensing requires a common-probe or ensemble risk
functional.  With nuisance prior \(\bm F_{\rm prior}\), all three use the efficient
Fisher matrix
\begin{equation}
 \bm F_{\rm eff}
 =
 \bm F_{\theta\theta}
 -
 \bm F_{\theta\nu}
 (\bm F_{\nu\nu}+\bm F_{\rm prior})^+
 \bm F_{\nu\theta},
 \label{eq:ensembleSchur}
\end{equation}
together with its support condition.

For statistically independent complex records, let \(M_r\) be the number of
records acquired at frequency \(\omega_r\), let \(\Delta t_r\) be the duration
of one record, and let \(\bm F_{\rm rec}^{(r)}\) denote the per-record Fisher
matrix.  Then
\begin{equation}
 \bm F_{\rm data}^{\rm tot}
 =
 \sum_r M_r\bm F_{\rm rec}^{(r)},\qquad
 \sum_r M_r\le M_{\rm tot},\qquad
 \sum_r M_rP_r\Delta t_r\le E_{\rm src}.
 \label{eq:frequencyAllocation}
\end{equation}
The last constraint is a sequential source-energy budget.  Equal
\(\Delta t_r\) gives a fixed-duration design.  For simultaneous multitone
records, the same sum applies to statistically independent or jointly whitened
frequency bins; otherwise the likelihood retains their joint covariance.

\suppsection{Reachable Fisher regions under resource constraints}
\suppsubsection{General positive-semidefinite resource theorem}
\label{sec:generalresource}

Let $\bm R\succeq0$, $R_0\ge0$, and impose
\begin{equation}
 \bm s^\dagger\bm R\bm s\le R_0.
 \label{eq:generalResource}
\end{equation}
Define $\bm T=\bm\Sigma_c^{-1/2}(\partial_\epsilon\bm S)$.

\textbf{Theorem S1 (single quadratic resource).}  The optimum is finite if and
only if
\begin{equation}
 \ker\bm R\subseteq\ker\bm T.
 \label{eq:finitecondition}
\end{equation}
When Eq.~\eqref{eq:finitecondition} holds,
\begin{equation}
 F_{\mu,R}^{\star}=2R_0\norm{\bm T(\bm R^+)^{1/2}}_2^2.
 \label{eq:genproofresult}
\end{equation}

\emph{Proof.}  If $\bm n\in\ker\bm R$ but $\bm T\bm n\ne0$, then
$\bm s=t\bm n$ has zero declared cost and Fisher information proportional to
$|t|^2$, so the problem is unbounded.  Conversely, under
Eq.~\eqref{eq:finitecondition}, the null-space component is informationless and
may be removed.  Every representative on
$\operatorname{supp}\bm R=(\ker\bm R)^\perp$ can be written as
\begin{equation}
 \bm s=(\bm R^+)^{1/2}\bm x,
 \qquad \bm x\in\operatorname{supp}\bm R,
 \qquad \norm{\bm x}^2\le R_0.
\end{equation}
The singular-value variational principle gives Eq.~\eqref{eq:genproofresult}.
If the largest singular value is positive, a unit principal right singular
vector $\bm v_1$ lies in $\operatorname{supp}\bm R$, and one optimum is
\begin{equation}
 \bm s_\star=\sqrt{R_0}\,(\bm R^+)^{1/2}\bm v_1.
 \label{eq:optinputgeneral}
\end{equation}
If the largest singular value vanishes, every feasible input has zero
mean-response information and $\bm s=0$ is the minimum-norm optimum.
\hfill$\square$

For a semidefinite metric, the optimization is performed on
\(\operatorname{supp}\bm R\) after checking
Eq.~\eqref{eq:finitecondition}.  An informative direction in \(\ker\bm R\)
makes the stated problem unbounded.  An informationless null direction can be
quotiented out.  Equation~\eqref{eq:genproofresult} is the
resource-metric form of the maximum-information-state eigenproblem
~\cite{Bouchet2021} and provides the input-space result for comparing physical
resources.

\suppsubsection{Reachable mean-FI region under an internal-resource budget}
\label{sec:coverage}

Let $\bm a=\bm G\bm B\bm s$ and $U=\bm a^\dagger\bm Q\bm a$, with
$\bm Q\succ0$ calibrated at the nominal point and held fixed in the resource
comparison.  If the internal metric depended on the target parameter,
differentiating the resource would add
\begin{equation}
 \dot U=2\operatorname{Re}(\bm a^\dagger\bm Q_0\dot{\bm a})
 +\bm a^\dagger\dot{\bm Q}\bm a.
 \label{eq:movingResourceMetric}
\end{equation}
Introduce
\begin{equation}
 \bm z=\bm Q^{1/2}\bm a=\bm X\bm s,
 \qquad \bm X=\bm Q^{1/2}\bm G\bm B,
 \qquad \bm\Pi_X=\bm X\bm X^+.
 \label{eq:Xz}
\end{equation}
The internal fields $\bm z$ that the ports can prepare form $\operatorname{ran}\bm X$.

For parameters that enter through
\(\bm V_\alpha=\partial_\alpha\bm H_{\rm eff}\), define the downstream
noise-whitened maps
\begin{equation}
 \bm A_\alpha=
 \bm\Sigma_c^{-1/2}\bm C\bm G
 \bm V_\alpha\bm Q^{-1/2}.
 \label{eq:parameterMapsSupp}
\end{equation}
An ensemble of coherent probes has the internal design moment
\begin{equation}
 \bm\Xi_z=\sum_k n_k\bm z_k\bm z_k^\dagger\succeq0,
 \qquad \Tr\bm\Xi_z=\sum_kn_kU_k.
 \label{eq:internalMomentSupp}
\end{equation}

\textbf{Theorem S2 (reachable mean-FI region at fixed $U$).}
Allow an ensemble containing up to \(\rank\bm X\) independently
weighted coherent probe settings and impose the aggregate
internal-resource budget \(\Tr\bm\Xi_z\le U_0\).  The attainable mean-FI
matrices are exactly
\begin{equation}
 \begin{split}
 \mathfrak R_{\mu,U}={}\Big\{&
 [\,2\operatorname{Re}\Tr(
 \bm\Xi_z\bm A_\alpha^\dagger\bm A_\beta)\,]_{\alpha\beta}:\\[-2pt]
 &\bm\Xi_z\succeq0,\quad
 \bm\Pi_X\bm\Xi_z\bm\Pi_X=\bm\Xi_z,\quad
 \Tr\bm\Xi_z\le U_0\Big\}.
 \end{split}
 \label{eq:FisherRegionSupp}
\end{equation}

\emph{Proof.}
 Every probe gives \(\bm z_k=\bm X\bm s_k\), hence a positive-semidefinite
 moment supported on \(\ran\bm X\), with trace equal to the allocated resource.
Equation~\eqref{eq:ensembleFIMSupp} then gives
\begin{equation}
 [\bm F_\mu]_{\alpha\beta}
 =2\operatorname{Re}\Tr(
 \bm\Xi_z\bm A_\alpha^\dagger\bm A_\beta).
 \label{eq:FisherRegionForwardSupp}
\end{equation}
Conversely, decompose an admissible moment as
\(\bm\Xi_z=\sum_{\ell=1}^r\bm z_\ell\bm z_\ell^\dagger\), where
\(r\le\rank\bm X\).  Its support condition permits the inputs
\(\bm s_\ell=\bm X^+\bm z_\ell\), which realize the same moment.
\hfill\(\square\)

Equation~\eqref{eq:FisherRegionSupp} is a linear image of the
positive-semidefinite trace ball on the reachable subspace.  Two preparation
maps with the same projector \(\bm\Pi_X\) have the same mean-FI region when
\(\bm Q\), the maps \(\{\bm A_\alpha\}\), and the monitored covariance are the same.
If \(\ran\bm X_1\subseteq\ran\bm X_2\), then
\(\mathfrak R_{\mu,U}^{(1)}\subseteq\mathfrak R_{\mu,U}^{(2)}\).
These relations order the local mean-Fisher design regions.  Ordering the full
likelihoods additionally requires the covariance scores.

For a fixed record count, one also has
\(\rank\bm\Xi_z\le N_{\rm rec}\); this is inactive once
\(N_{\rm rec}\ge\rank\bm X\).  Under the protocol of
Sec.~\ref{sec:completeProtocol}, equality or inclusion extends to the full
Fisher designs.

\textbf{Corollary S2.1 (scalar reachable-projector optimum).}
For an arbitrary finite-rank scalar parameter tangent $\bm V$ and
$U_0\ge0$,
\begin{equation}
 F_{\mu,U}^{\star}=2U_0\norm{\bm A\bm\Pi_X}_2^2,
 \quad
 \bm A=\bm\Sigma_c^{-1/2}\bm C\bm G\bm V\bm Q^{-1/2}.
 \label{eq:projectortheoremsupp}
\end{equation}

\emph{Proof.}  Equations~\eqref{eq:dSsupp} and~\eqref{eq:Xz} imply
\begin{equation}
 \bm\Sigma_c^{-1/2}(\partial_\epsilon\bm S)\bm s
 =\ii\bm A\bm z.
\end{equation}
Therefore
\begin{equation}
 F_{\mu,U}^{\star}
 =\max_{\bm z\in\ran\bm X,\,\norm{\bm z}^2\le U_0}2\norm{\bm A\bm z}^2
 =2U_0\norm{\bm A\bm\Pi_X}_2^2.
\end{equation}
\hfill$\square$

\textbf{Corollary S2.2 (minimum source cost of an internal Fisher design).}
Let \(\bm R\succ0\) and
\(\bm Y=\bm X\bm R^{-1/2}\).  The minimum aggregate source-resource cost required
to realize an admissible internal moment is
\begin{equation}
 P_{\min}(\bm\Xi_z)
 =\Tr\!\left[
 \bm Y^+\bm\Xi_z(\bm Y^+)^\dagger
 \right].
 \label{eq:internalDesignSourcePrice}
\end{equation}
Indeed, for each component \(\bm z_\ell\), the substitution
\(\bm u_\ell=\bm R^{1/2}\bm s_\ell\) reduces the minimum-source
preparation problem to
\(\min_{\bm Y\bm u_\ell=\bm z_\ell}\norm{\bm u_\ell}^2\), whose
minimum-norm solution is \(\bm Y^+\bm z_\ell\).  Summing over a
factorization of \(\bm\Xi_z\) gives
Eq.~\eqref{eq:internalDesignSourcePrice}, independently of the chosen
factorization.  Preparation maps with the same reachable projector can have
identical mean-FI regions at the same $U$ but different
preparation-cost functions.

Equivalently, with
\(\bm W_0=\bm Y\bm Y^\dagger\),
\[
 P_{\min}(\bm\Xi_z)=\Tr(\bm W_0^+\bm\Xi_z)
\]
for moments supported on \(\ran\bm Y\).  Let \(\mathcal L\) denote the
linear map from internal design moments to Fisher matrices in
Eq.~\eqref{eq:FisherRegionSupp}.  The minimum preparation cost of a Fisher-region
point is the minimum of this expression over all internal moments that produce
that Fisher matrix:
\begin{equation}
 P_{\min}(\bm F_\mu)=
 \min_{\substack{\bm\Xi_z\succeq0,\,
                  \bm\Pi_X\bm\Xi_z\bm\Pi_X=\bm\Xi_z,\,
                  \Tr\bm\Xi_z\le U_0\\
                  \mathcal L(\bm\Xi_z)=\bm F_\mu}}
 \Tr(\bm W_0^+\bm\Xi_z).
 \label{eq:FisherPointSourcePrice}
\end{equation}
Different internal moments can represent the same Fisher matrix, and their
minimum source
cost depends on both the eigenvectors and eigenvalues of the full Gramian.
Figure 1 of the main text illustrates this separation in a stable
four-mode, two-parameter example with reachable rank \(2\). Its boundary
points maximize \((1-q)F_{\mu,11}+qF_{\mu,22}\); the second panel compares
the minimum source costs of two preparation maps with the same
\(\bm\Pi_X\).

If $\bm r_1$ is a principal right singular vector of
$\bm A\bm\Pi_X$ with nonzero singular value, an optimal state and its
minimum-Euclidean-norm port input are
\begin{equation}
 \bm z_\star=\sqrt{U_0}\,\bm r_1,
 \qquad
 \bm a_\star=\bm Q^{-1/2}\bm z_\star,
 \qquad
 \bm s_\star=\bm X^+\bm z_\star.
 \label{eq:optimalStates}
\end{equation}
If the singular value vanishes, the reachable mean response carries no target
information.  Small nonzero singular values of $\bm X$ increase the source
cost of preparing $\bm z_\star$ without changing the fixed-\(U\) optimum.

For a rank-one task $\bm V_j=\bm v_j\bm w_j^\dagger$, define
\begin{equation}
 \Ecal_j=\norm{\bm\Sigma_c^{-1/2}\bm C\bm G\bm v_j}^2,
 \qquad \widetilde{\bm w}_j=\bm Q^{-1/2}\bm w_j.
\end{equation}
Then
\begin{equation}
 F_{\mu,U,B,j}^{\star}=2U_0\Ecal_j\,
 \widetilde{\bm w}_j^\dagger\bm\Pi_X\widetilde{\bm w}_j.
 \label{eq:fixedUrankone}
\end{equation}
For the Euclidean source metric \(\bm R=\bm I\), introduce
\begin{equation}
 \alpha_j=
 \frac{\widetilde{\bm w}_j^\dagger\bm\Pi_X\widetilde{\bm w}_j}
 {\norm{\widetilde{\bm w}_j}^2},
 \qquad
 \beta_j=
 \frac{\norm{\bm X^\dagger\widetilde{\bm w}_j}^2}
 {\widetilde{\bm w}_j^\dagger\bm\Pi_X\widetilde{\bm w}_j},
 \label{eq:alphaBetaSupp}
\end{equation}
when the projector overlap in the denominator is nonzero.  The
source-resource optimum then factorizes as
\begin{equation}
 F_{\mu,P,B,j}^{\star}
 =2P_0\Ecal_j\norm{\bm X^\dagger\widetilde{\bm w}_j}^2
 =2P_0\Ecal_j\norm{\widetilde{\bm w}_j}^2\alpha_j\beta_j.
 \label{eq:fixedPrankone}
\end{equation}
The factor \(\alpha_j\) is the normalized squared projection of the weighted
task direction onto the reachable subspace.  The factor \(\beta_j\) measures
preparation buildup on that component.  If
\(\alpha_j=0\), then \(\bm X^\dagger\widetilde{\bm w}_j=0\), and both
the fixed-source and fixed-\(U\) mean-FI optima
vanish; \(\beta_j\) need not be assigned when \(\alpha_j=0\).
The ideal full-state drive has \(\bm\Pi_X=\bm I\).  When
\(U_0>0\), \(\widetilde{\bm w}_j\ne0\), and
\(F_{\mu,U,{\rm full},j}^{\star}
=2U_0\Ecal_j\norm{\widetilde{\bm w}_j}^2>0\), the corresponding retention is
\begin{equation}
 \Ccal_j\equiv
 \frac{F_{\mu,U,B,j}^{\star}}{F_{\mu,U,{\rm full},j}^{\star}}
 =\frac{\widetilde{\bm w}_j^\dagger\bm\Pi_X\widetilde{\bm w}_j}
 {\norm{\widetilde{\bm w}_j}^2}.
 \label{eq:coverageSupp}
\end{equation}
The ratio satisfies \(0\le\Ccal_j\le1\), and
\(\Ccal_j=\cos^2\theta_j\) is the squared cosine of the principal angle between
the weighted task direction and the internally reachable subspace.  It is also
a leverage score of the reachable projector~\cite{Mahoney2011,Drineas2012}.
In passive systems a
consistently normalized Wigner--Smith quadratic form can define the internal
dwell metric~\cite{Wigner1955,Smith1960,MahauxWeidenmuller1969}; each example
below uses the calibrated \(\bm Q\) specified for that model.

\suppsubsection{Retention budget, subspace geometry, and probe order}
\label{sec:geometry}

Let $\{\widetilde{\bm w}_j\}_{j=1}^N$ be a complete orthonormal basis of the
internal task space, and assume that each full-state-drive Fisher denominator is
nonzero.  Equation~\eqref{eq:coverageSupp} gives
\begin{equation}
 \sum_{j=1}^N\Ccal_j=\Tr\bm\Pi_X=\rank\bm X\equiv m.
 \label{eq:sumruleSupp}
\end{equation}
For any subset $\mathcal J$ with projector
$\bm P_{\mathcal J}=\sum_{j\in\mathcal J}
\widetilde{\bm w}_j\widetilde{\bm w}_j^\dagger$, the corresponding partial
sum is
\begin{equation}
 \sum_{j\in\mathcal J}\Ccal_j
 =\Tr(\bm P_{\mathcal J}\bm\Pi_X).
 \label{eq:partialSumruleSupp}
\end{equation}
Only the complete basis reduces this expression to $\rank\bm X$.
For a fixed $\bm B$, this is the diagonal vector of a fixed projector.
Adjusting the complex input amplitudes moves the prepared state inside
$\ran\bm X$ and leaves $\bm\Pi_X$ fixed.  Allowing the rank-\(m\) internal
control subspace itself to be redesigned makes the full hypersimplex
\begin{equation}
 \Delta(N,m)=\left\{\bm c\in[0,1]^N:\sum_jc_j=m\right\}
 \label{eq:hypersimplex}
\end{equation}
reachable.  Selecting $m$ local physical ports from a fixed lattice
generally produces a discrete, nonconvex subset.

For two devices on the same resource-orthonormal task basis and with equal
reachable rank, if
\begin{equation}
 \Ccal_j^{(A)}\ge\Ccal_j^{(B)}\qquad\text{for every }j,
 \label{eq:coverageParetoAssumption}
\end{equation}
then Eq.~\eqref{eq:sumruleSupp} implies
\begin{equation}
 \Ccal_j^{(A)}=\Ccal_j^{(B)}\qquad\text{for every }j.
 \label{eq:coverageNoPareto}
\end{equation}
At fixed rank, a gain for one task is balanced elsewhere unless the profiles
coincide.  Emission and output statistics set the absolute Fisher scale.

For arbitrary rank-$m$ subspace design, the Schur--Horn theorem
~\cite{Schur1923,Horn1954,HornJohnson2013} makes every vector in
Eq.~\eqref{eq:hypersimplex} attainable.  Hence
\begin{equation}
 \max_{\rank\bm\Pi=m}\min_j\langle j|\bm\Pi|j\rangle=\frac{m}{N},
 \label{eq:nonlocalMinimax}
\end{equation}
with equality achieved by any $m$ columns of the discrete Fourier matrix.  For
a prior matrix $\bm M=\sum_jp_j|j\rangle\langle j|$, the average optimum is
\begin{equation}
 \max_{\rank\bm\Pi=m}\Tr(\bm M\bm\Pi)=\sum_{k=1}^m\lambda_k(\bm M).
 \label{eq:KyFan}
\end{equation}
The value \(m/N\) is the nonlocal reference for the local-port bound below.

For an \(r\)-dimensional parameter subspace with
\(\bm W^\dagger\bm W=\bm I_r\), the eigenvalues of
\(\bm W^\dagger\bm\Pi_X\bm W\) are
\(\cos^2\theta_k\), the principal-angle spectrum between the parameter and
reachable subspaces~\cite{BjorckGolub1973}.  This spectrum provides the finite-rank
extension of the single-task quotient.

For localized port responses, the exact feed-forward law can be compared with
a more general bound.  Normalize the independent columns of \(\bm X\) to obtain
\(\overline{\bm X}\), and let
\(\bm G_p=\overline{\bm X}^{\dagger}\overline{\bm X}\).  If
\[
 |\overline X_{j\ell}|\le c\,e^{-d(j,p_\ell)/\xi},
 \qquad \lambda_{\min}(\bm G_p)\ge g_0>0,
\]
then
\begin{equation}
 \Ccal_j
 \le \frac{mc^2}{g_0}
 e^{-2d(j,\mathcal P)/\xi},
 \qquad d(j,\mathcal P)=\min_\ell d(j,p_\ell).
 \label{eq:conditionalLocalityBound}
\end{equation}
This follows directly from
\(\Ccal_j=\bm r_j\bm G_p^{-1}\bm r_j^\dagger\), where
\(\bm r_j=\langle j|\overline{\bm X}\).
The exact chain result below is stronger because its nested response columns
admit an explicit orthogonal decomposition.

The exact profile is also a reference for rank-preserving weak departures
from the feed-forward model.  Let \(\bm X=\bm X_0+\Delta\bm X\), with
\(\rank\bm X=\rank\bm X_0=m\), and set
\(\epsilon=\norm{\Delta\bm X}_2\).  If
\(\epsilon<\sigma_m(\bm X_0)\), then for every nonzero task direction
\(\widetilde{\bm w}\),
\begin{align}
 \left|
 \frac{\widetilde{\bm w}^\dagger\bm\Pi_X\widetilde{\bm w}}
 {\norm{\widetilde{\bm w}}^2}
 -
 \frac{\widetilde{\bm w}^\dagger\bm\Pi_{X_0}\widetilde{\bm w}}
 {\norm{\widetilde{\bm w}}^2}
 \right|
 &\le \norm{\bm\Pi_X-\bm\Pi_{X_0}}_2,\nonumber\\
 \norm{\bm\Pi_X-\bm\Pi_{X_0}}_2
 &\le \min\!\left\{1,
 \frac{\epsilon}{\sigma_m(\bm X_0)-\epsilon}\right\}.
 \label{eq:projectorPerturbationSupp}
\end{align}
The first inequality is the Rayleigh-quotient bound.  For the second, choose
the minimum-norm coefficient vector for a unit vector in \(\ran\bm X\).
Its norm is at most \(1/\sigma_m(\bm X)\), and Weyl's inequality gives
\(\sigma_m(\bm X)\ge\sigma_m(\bm X_0)-\epsilon\).  Projecting the resulting
representation onto \((\ran\bm X_0)^\perp\) yields the stated principal-angle
bound; equal ranks identify it with the projector distance
~\cite{Wedin1972,BjorckGolub1973}.  Reverse hopping, detuning, and component
nonuniformity enter this result through the induced \(\Delta\bm X\).

Multiprobe acquisition budgets are derived in Sec.~\ref{sec:completeProtocol};
the exact common-probe bound is derived in Sec.~\ref{sec:covNuisance}.

\suppsection{Joint source--internal-resource optimization}
\suppsubsection{Two-resource problem}
\label{sec:joint}

For a finite-rank parameter tangent $\bm V$, define
\begin{equation}
 \bm A=\bm\Sigma_c^{-1/2}\bm C\bm G\bm V\bm Q^{-1/2},
 \qquad
 \bm H=\bm X^\dagger\bm X,
 \qquad
 \bm K=\bm X^\dagger\bm A^\dagger\bm A\bm X.
 \label{eq:AHK}
\end{equation}
The joint resource problem is
\begin{align}
 \max_{\bm s}\quad &2\bm s^\dagger\bm K\bm s,\nonumber\\
 \text{subject to}\quad &\bm s^\dagger\bm R\bm s\le P_0,
 \qquad
 \bm s^\dagger\bm H\bm s\le U_0,
 \label{eq:jointPrimal}
\end{align}
where $\bm R,\bm H,\bm K\succeq0$.

\textbf{Theorem S3 (joint homogeneous resource duality).}  Let
\(\mathcal{N}_0=\ker\bm R\cap\ker\bm H\).  The optimum of
Eq.~\eqref{eq:jointPrimal} is finite if and only if
\begin{equation}
 \ker\bm R\cap\ker\bm H\subseteq\ker\bm K.
 \label{eq:jointFinite}
\end{equation}
If $P_0,U_0>0$ and Eq.~\eqref{eq:jointFinite} holds, then
\begin{equation}
 F_{\mu,P,U}^{\star}=2\min_{\lambda_P,\lambda_U\ge0}
 (\lambda_P P_0+\lambda_U U_0)
 \quad\text{subject to}\quad
 \bm K\preceq\lambda_P\bm R+\lambda_U\bm H.
 \label{eq:jointDualSupp}
\end{equation}
There exists an optimal vector $\bm s_\star$, equivalently a rank-one optimal
lift $\bm Z_\star=\bm s_\star\bm s_\star^\dagger$.

\emph{Finiteness.}  A vector in
$\ker\bm R\cap\ker\bm H$ with positive $\bm K$ quadratic form can be scaled
without cost, making the objective unbounded.  On the orthogonal complement of
this common kernel, $\bm R+\bm H$ is positive definite and
$\bm K\preceq c(\bm R+\bm H)$ for some finite $c$.  In the physical
factorization, $\ker\bm H=\ker\bm X\subseteq\ker\bm K$.

\emph{Strong duality.}  Quotient out $\mathcal{N}_0$, or equivalently restrict
the problem to $\mathcal{N}_0^\perp$.  On this subspace, $\bm R+\bm H$ is
positive definite and the feasible set is compact.  Moreover, $\bm s=\bm0$
strictly satisfies both inequalities when $P_0,U_0>0$.  The complex homogeneous
QCQP is therefore in the setting of Theorem~2.2 of Beck and
Eldar~\cite{BeckEldar2006}: its Lagrange dual is exact and the optimum is attained by
a vector.  In lifted form,
\begin{align}
 \max_{\bm Z\succeq0}\quad &2\Tr(\bm K\bm Z),\nonumber\\
 \text{subject to}\quad &\Tr(\bm R\bm Z)\le P_0,
 \qquad \Tr(\bm H\bm Z)\le U_0.
 \label{eq:jointSDP}
\end{align}
The Lagrange dual is Eq.~\eqref{eq:jointDualSupp}, and the vector optimum gives
the rank-one lift stated above.  This quotient formulation is required when the
two resource matrices have a common kernel.

\emph{KKT recovery.}  Let $(\lambda_P^\star,\lambda_U^\star)$ solve
Eq.~\eqref{eq:jointDualSupp}, and define the dual slack
\begin{equation}
 \bm M_\star=\lambda_P^\star\bm R+\lambda_U^\star\bm H-\bm K\succeq0.
\end{equation}
An optimal vector satisfies
\begin{align}
 \bm M_\star\bm s_\star&=0,\label{eq:KKTnull}\\
 \lambda_P^\star(\bm s_\star^\dagger\bm R\bm s_\star-P_0)&=0,\nonumber\\
 \lambda_U^\star(\bm s_\star^\dagger\bm H\bm s_\star-U_0)&=0.
 \label{eq:KKTcomp}
\end{align}
When $\ker\bm M_\star$ is one dimensional, its basis vector is scaled to the
active constraint.  In a degenerate mixed region, write
$\bm s=\bm N_\star\bm y$ and solve
\(\bm y^\dagger\bm N_\star^\dagger
[\bm H-(U_0/P_0)\bm R]\bm N_\star\bm y=0\) in the reduced coordinates before
scaling.

For a zero budget, the problem is first restricted to the corresponding
resource kernel.

The fixed-source and fixed-\(U\) limits are
\begin{align}
 F_{\mu,P}^\star&=2P_0\lambda_{\max}(\bm K,\bm R),
 \label{eq:fixedPgeneral}\\
 F_{\mu,U}^\star&=2U_0\lambda_{\max}(\bm K,\bm H)
 =2U_0\norm{\bm A\bm\Pi_X}_2^2,
 \label{eq:fixedUgeneral}
\end{align}
with the corresponding support conditions for singular metrics.

\suppsubsubsection{Resource filtering of the mean-FI kernel}

For the physical factorization
\(\bm H=\bm X^\dagger\bm X\) and
\(\bm K=\bm X^\dagger\bm A^\dagger\bm A\bm X\), set
\(\lambda_P=q\) and \(\lambda_U=q\tau\), with \(q>0\) and \(\tau\ge0\).
For fixed \(\tau\), the smallest admissible scale is
\begin{equation}
 \varphi(\tau)=
 \lambda_{\max}\!\left[
 \bm A\bm X(\bm R+\tau\bm H)^+
 \bm X^\dagger\bm A^\dagger\right],
 \label{eq:rayPhi}
\end{equation}
provided
\(\ker(\bm R+\tau\bm H)\subseteq\ker\bm K\).
The joint optimum is
\begin{equation}
 \frac{F_{\mu,P,U}^{\star}}{2}
 =
 \min\!\left\{
 \inf_{\tau\ge0}(P_0+\tau U_0)\varphi(\tau),
 U_0\norm{\bm A\bm\Pi_X}_2^2
 \right\},
 \label{eq:jointSpectralEnvelopeSupp}
\end{equation}
where the second term is the pure internal-resource endpoint.

When \(\bm R\succ0\) on the admitted input space, define
\(\bm X_R=\bm X\bm R^{-1/2}\) and
\(\bm W_0=\bm X_R\bm X_R^\dagger\).  Direct SVD gives
\begin{equation}
 \bm W_\tau
 =
 \bm X(\bm R+\tau\bm X^\dagger\bm X)^{-1}\bm X^\dagger
 =
 \bm W_0(\bm I+\tau\bm W_0)^{-1}.
 \label{eq:WtauIdentity}
\end{equation}
Each nonzero preparation eigenvalue is filtered as
\begin{equation}
 w_i\longmapsto\frac{w_i}{1+\tau w_i},
 \qquad
 \bm W_{\tau=0}=\bm W_0,\qquad
 \lim_{\tau\to\infty}\tau\bm W_\tau=\bm\Pi_X.
 \label{eq:spectralClipping}
\end{equation}
At the internal-resource endpoint, the nonzero preparation singular values
drop out and only their span remains.  For the one-port chain,
\(w_N=\gamma^{-2}S_N\), with
\(S_N=\sum_{r=0}^{N-1}\rho^{2r}\), so
\begin{equation}
 \phi_N(\widehat\tau)
 =\frac{w_{\tau,N}}{w_N}
 =\frac{1}{1+\widehat\tau S_N},
 \qquad \widehat\tau=\tau/\gamma^2.
 \label{eq:chainSpectralFilterSupp}
\end{equation}

Let \(w_{\max}\) and \(w_{\min,+}\) be the extreme positive eigenvalues of
\(\bm W_0\).  A mixed solution can occur only when
\begin{equation}
 w_{\min,+}<\frac{U_0}{P_0}<w_{\max},
 \label{eq:mixedSpectralBand}
\end{equation}
although the orientation of \(\bm A\) may narrow this interval.
For singular \(\bm R\), Eq.~\eqref{eq:jointDualSupp} remains the primary
formulation: zero-source-cost directions are retained when they consume
internal resource, and only the common zero-cost information-null kernel is
removed.

\suppsubsubsection{Resource regimes and marginal values}

Let $t=U_0/P_0$.  If a fixed-source maximizer satisfies the internal constraint,
$\lambda_U^\star=0$; if a fixed-\(U\) maximizer satisfies the
source constraint, $\lambda_P^\star=0$; otherwise both constraints are active.
For finite positive one-resource optima, let
\begin{equation}
 \mathcal E_P=\{\bm v:\bm K\bm v=\rho_P\bm R\bm v,\ \bm v^\dagger\bm R\bm v=1\},
\end{equation}
where $\rho_P$ is the largest supported generalized eigenvalue, and define
\begin{equation}
 h_P^{\min}=\min_{\bm v\in\mathcal E_P}\bm v^\dagger\bm H\bm v.
\end{equation}
Then $t\ge h_P^{\min}$ is the source-limited condition.  Similarly, with
\begin{equation}
 \mathcal E_U=\{\bm v:\bm K\bm v=\rho_U\bm H\bm v,\ \bm v^\dagger\bm H\bm v=1\},
 \qquad
 r_U^{\min}=\min_{\bm v\in\mathcal E_U}\bm v^\dagger\bm R\bm v,
\end{equation}
$t\le1/r_U^{\min}$ is the internal-resource-limited condition, with
$1/r_U^{\min}=+\infty$ for $r_U^{\min}=0$.  Away from degenerate boundaries,
the envelope theorem gives
\begin{equation}
 \frac{\partial F_{\mu,P,U}^\star}{\partial P_0}=2\lambda_P^\star,
 \qquad
 \frac{\partial F_{\mu,P,U}^\star}{\partial U_0}=2\lambda_U^\star.
 \label{eq:marginalValues}
\end{equation}
The quantities $2\lambda_P^\star$ and $2\lambda_U^\star$ are the marginal mean
FI per unit source and internal resource.

\suppsubsubsection{Four-mode joint frontier}

\begin{figure}[t]
 \centering
 \includegraphics[width=0.52\linewidth]{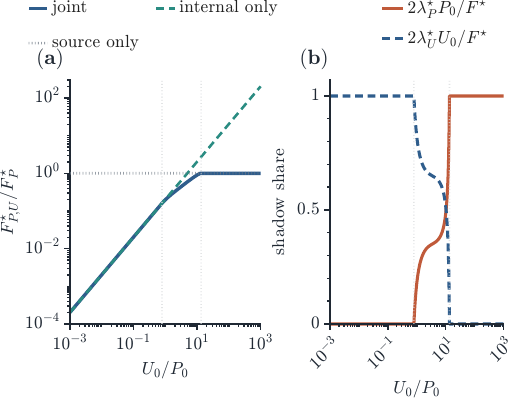}
 \caption{\textbf{Four-mode source--internal-resource frontier.}
 (a) Joint optimum normalized by the source-only optimum, together with the
 two one-resource endpoints.  (b) Normalized budget-weighted shadow-price
 contributions from the source and internal resources.  Both are nonzero in
 the mixed interval and sum to unity.}
 \label{fig:jointFrontierSupp}
\end{figure}

The frontier in Fig.~\ref{fig:jointFrontierSupp} uses a stable four-mode
feed-forward generator and two
local drive ports, with
\(\bm T_+=\sum_{j=1}^{3}|j+1\rangle\langle j|\),
\begin{equation}
 \bm H_{\rm eff}=-\ii\bm I+1.35\bm T_+,
 \qquad
 \bm B=(|1\rangle,|3\rangle),
 \qquad \bm R=\bm Q=\bm I.
 \label{eq:v27FrontierModel}
\end{equation}
The monitored rows correspond to sites 1, 3, and 4, with covariance
\begin{equation}
 \bm\Sigma_c=
 \begin{pmatrix}
 1&0.18&0.08\\
 0.18&0.75&0.12\\
 0.08&0.12&1.15
 \end{pmatrix},
 \label{eq:v27FrontierCovariance}
\end{equation}
and the scalar target tangent is
\begin{equation}
 \bm V=|2\rangle\langle2|+
 \frac12\left[0.70|3\rangle\langle3|+
 0.45(|2\rangle\langle3|+|3\rangle\langle2|)\right].
 \label{eq:v27FrontierTarget}
\end{equation}
The leading source- and internal-resource-optimal directions differ.  Their
endpoint feasibility conditions give
\begin{equation}
 \frac{U_0}{P_0}\le\JointInternalMixedBoundary
 \quad(\text{internal limited}),
 \qquad
 \frac{U_0}{P_0}\ge\JointMixedSourceBoundary
 \quad(\text{source limited}).
 \label{eq:v27FrontierBoundaries}
\end{equation}
Both multipliers are positive between these boundaries.  The rank-one KKT
solution, two-variable dual, and spectral envelope agree within
\(\ValVTwentySevenJointFrontier\); the normalized shadow-price contributions
sum to unity along the frontier.

\suppsection{Local-port laws and resource crossover}
\suppsubsection{Local-port retention in a strictly feed-forward chain}
\label{sec:unidirectional}

For $\gamma>0$ and real $\kappa\ge0$, consider
\begin{equation}
 \bm H_J=-\ii\gamma\bm I+\kappa\bm T_+,
 \qquad
 \bm T_+=\sum_{j=1}^{N-1}|j+1\rangle\langle j|,
 \qquad \rho=\kappa/\gamma.
 \label{eq:HJSupp}
\end{equation}
All eigenvalues equal $-\ii\gamma$, so the system remains stable for every
$\kappa$ and every $N$.  At $\omega=0$,
\begin{equation}
 \bm G_J=(\bm H_J)^{-1}
 =\frac{\ii}{\gamma}\sum_{r=0}^{N-1}(-\ii\rho)^r\bm T_+^r.
 \label{eq:GJseries}
\end{equation}
The ratio \(\rho\) compares forward coupling with loss.
The Green column produced by a port at $p$ has support on $j\ge p$ and
geometric magnitude
\begin{equation}
 |\langle j|\bm G_J|p\rangle|=\gamma^{-1}\rho^{j-p},
 \qquad j\ge p.
 \label{eq:GJcolumn}
\end{equation}
We take $\bm Q=\bm I$, local tasks
$\bm V_j=|j\rangle\langle j|$, full-state monitoring, and white covariance.
Every projector quotient is then the corresponding Fisher retention.  A
positive diagonal $\bm Q$ preserves the segment decomposition and weights its
geometric profile by the calibrated resource.

Let the ordered port sites be
\begin{equation}
 1=p_1<p_2<\cdots<p_m,
 \qquad p_{m+1}=N+1,
 \qquad L_\ell=p_{\ell+1}-p_\ell.
 \label{eq:portSegments}
\end{equation}
For $\rho>0$, the choice \(p_1=1\) makes every site reachable.  At
$\rho=0$, only the port sites are reachable.  If $p_1>1$, all tasks $j<p_1$
have zero retention and the task-specific worst-site optimum is zero.

\textbf{Theorem S4 (segment retention law).}  For $\rho>0$ and the ports in Eq.~\eqref{eq:portSegments},
\begin{equation}
 \Ccal_j=\frac{\rho^{2(j-p_\ell)}}{S_{L_\ell}},
 \qquad
 S_L=\sum_{r=0}^{L-1}\rho^{2r},
 \qquad p_\ell\le j<p_{\ell+1}.
 \label{eq:segmentLawSupp}
\end{equation}
The same decomposition gives
\begin{equation}
 \sum_{j=p_\ell}^{p_{\ell+1}-1}\Ccal_j=1
 \label{eq:segmentBudget}
\end{equation}
for every segment.
At $\rho=0$, Eq.~\eqref{eq:segmentLawSupp} is replaced by its direct
projector limit: \(\Ccal_j=1\) at a port site and \(\Ccal_j=0\) elsewhere.

\emph{Proof.}  The columns $\bm g_p=\bm G_J|p\rangle$ are nested.  The
component of $\bm g_{p_\ell}$ orthogonal to later-port columns is supported on
$p_\ell\le j<p_{\ell+1}$, with normalized vector
\begin{equation}
 |\bm u_\ell\rangle=\frac{1}{\sqrt{S_{L_\ell}}}
 \sum_{r=0}^{L_\ell-1}e^{\ii\varphi_r}\rho^r|p_\ell+r\rangle,
 \label{eq:segmentBasis}
\end{equation}
where the phases follow Eq.~\eqref{eq:GJseries}.  These vectors have disjoint
support, so
\begin{equation}
 \bm\Pi_X=\sum_{\ell=1}^m|\bm u_\ell\rangle\langle\bm u_\ell|,
\end{equation}
whose diagonal is Eq.~\eqref{eq:segmentLawSupp}.  Equation~\eqref{eq:segmentBudget} follows immediately.

For $\rho>1$, the weakest task in a segment is its upstream site and equals
\begin{equation}
 c(L)=S_L^{-1}=\frac{\rho^2-1}{\rho^{2L}-1}.
 \label{eq:segmentWorst}
\end{equation}
The function $c(L)$ is strictly decreasing in $L$.

\textbf{Theorem S5 (optimal local-port max--min retention).}  For
\(\rho>0\), among all $m$ local-port placements with $p_1=1$, the maximum
worst-task retention is obtained when the segment lengths differ by at most
one, and
\begin{equation}
 \Ccal_{\min}^{\rm loc,\star}
 =\frac{\rho^2-1}{\rho^{2\lceil N/m\rceil}-1},\qquad \rho>1.
 \label{eq:localMaxMinSupp}
\end{equation}
For $\rho=1$,
\begin{equation}
 \Ccal_{\min}^{\rm loc,\star}=\frac{1}{\lceil N/m\rceil}.
 \label{eq:rhoOneLimit}
\end{equation}
For $0<\rho<1$, the weakest task is at the downstream end of each segment.  Balanced placement remains optimal and, with $L_\star=\lceil N/m\rceil$, gives
\begin{equation}
 \Ccal_{\min}^{\rm loc,\star}
 =\frac{\rho^{2(L_\star-1)}}{\sum_{r=0}^{L_\star-1}\rho^{2r}},
 \qquad 0<\rho<1.
 \label{eq:rhoBelowOne}
\end{equation}

\emph{Proof.}  The weakest retention in a segment of length $L$ is
\begin{equation}
 c_\rho(L)=
 \begin{cases}
 S_L^{-1}, & \rho>1,\\
 L^{-1}, & \rho=1,\\
 \rho^{2(L-1)}S_L^{-1}, & 0<\rho<1.
 \end{cases}
 \label{eq:segmentWorstAllRho}
\end{equation}
Each branch decreases with $L$; for $0<\rho<1$, setting $q=\rho^2$ gives
\begin{equation}
 \frac{c_\rho(L+1)}{c_\rho(L)}
 =q\frac{1-q^L}{1-q^{L+1}}<1.
 \label{eq:segmentWorstBelowOneMonotonic}
\end{equation}
Thus the worst retention is $c_\rho(L_{\max})$.  Since
$\sum_\ell L_\ell=N$, the smallest possible $L_{\max}$ is
$\lceil N/m\rceil$, attained by balanced segments. \hfill$\square$

The arbitrary-subspace reference of Eq.~\eqref{eq:nonlocalMinimax} is $m/N$.
The local-to-arbitrary-subspace ratio is
\begin{equation}
 \frac{\Ccal_{\min}^{\rm loc,\star}}{m/N}
 =\frac{N}{m}\frac{\rho^2-1}{\rho^{2\lceil N/m\rceil}-1}
 \asymp \frac{N}{m}(\rho^2-1)\rho^{-2N/m},
 \qquad N/m\to\infty.
 \label{eq:localityPenalty}
\end{equation}
For $\rho>1$, this ratio decays exponentially with $N/m$.  At $\rho=1$,
integer partitioning is the sole difference between the local and
arbitrary-subspace optima.

For \(\rho>1\) and \(0<\Ccal_0\le1\), the largest segment length compatible with $\Ccal_j\ge\Ccal_0$ is
\begin{equation}
 L_{\max}(\rho,\Ccal_0)=
 \left\lfloor\frac{\ln[1+(\rho^2-1)/\Ccal_0]}{2\ln\rho}\right\rfloor,
 \qquad \rho>1,
 \label{eq:LmaxSupp}
\end{equation}
provided $L_{\max}\ge1$.  Consequently,
\begin{equation}
 m\ge\left\lceil\frac{N}{L_{\max}}\right\rceil.
 \label{eq:portCountSupp}
\end{equation}
For fixed $\rho>1$ and fixed nonzero $\Ccal_0$, $L_{\max}$ is independent of
$N$, and the required number of local ports scales linearly with \(N\).

Reverse hopping breaks the nested-column decomposition, so the exact segment
law is special to the feed-forward limit.  Direct projector calculations
recover Eq.~\eqref{eq:segmentLawSupp} at zero reverse hopping to
\(\ValReverseHoppingChiZero\) and show how the sitewise retention is redistributed
away from that limit while the rank sum remains fixed.  The local
rank-preserving perturbation bound applies when
\(\norm{\Delta\bm X}_2<\sigma_m(\bm X_0)\).  The nonperturbative optima considered below
lie outside this domain and are obtained directly from the reachable projector.

\begin{figure}[!t]
 \centering
 \includegraphics[width=13.3cm]{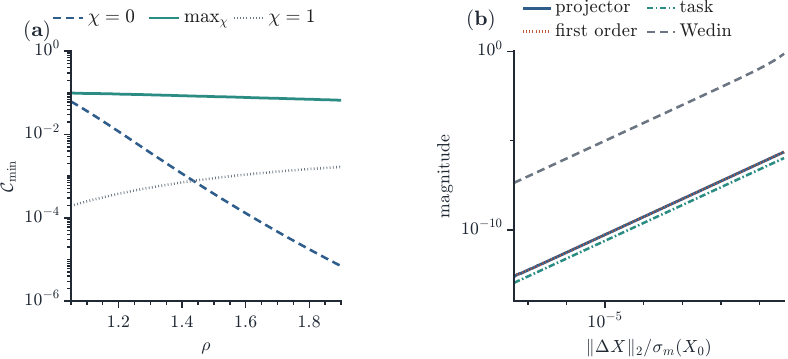}
 \caption{\textbf{Reverse-hopping interpolation and local subspace
 response.}
 (a) Worst-site retention at the feed-forward endpoint, after optimization
 over the stated \(\chi\) grid, and at the reciprocal endpoint.  The optimizer
 \(\chi^\star(\rho)\) is the ridge shown in Fig.~3(b) of the main text.
 (b) Within the rank-preserving neighborhood, the directly computed projector
 distance, the largest task-retention change, the first-order tangent
 \(\|(I-\Pi_0)\Delta X X_0^+\|_2\), and the Wedin bound.  The tangent follows
 the direct rotation, whereas the worst-case bound is much more conservative.}
 \label{fig:reverseNonperturbativeSupp}
\end{figure}

The reverse-hopping scan uses
\begin{equation}
 \bm H_\chi=-\ii\gamma\bm I+\kappa\bm T_+
             +\chi\kappa\bm T_-,\qquad \bm T_-=\bm T_+^{\mathsf T},
 \label{eq:reverseHoppingSupp}
\end{equation}
with \(\omega=0\), \(\gamma=1\), \(N=30\), and local ports
\(\{1,11,21\}\).  Thus \(\chi=0\) is the feed-forward chain and \(\chi=1\)
is reciprocal.  The main scan uses 61 uniformly spaced values of
\(\rho\in[1.05,1.90]\) and
\(\mathcal G_\chi=\{0,0.0025,\ldots,1\}\); the plotted optimizer is the grid
point at which the worst-site retention is maximal.
Representative profiles at \(\rho=1.15,1.35,1.55,1.85\) use 241 values of
\(\chi\).  The
maximizing value \(\chi^\star\) ranges from \(0.0475\) to \(0.4325\), as shown
in Fig.~\ref{fig:reverseNonperturbativeSupp}.  For \(\chi>0\), the hopping
matrix is similar to a real symmetric
tridiagonal matrix, whereas at \(\chi=0\) it is triangular; hence the poles
have imaginary part \(-\gamma\) throughout the scan.  The interpolation
follows how reverse hopping changes the drive-reachable subspace at fixed loss.

For one port, the retention profile is
\begin{equation}
 \Ccal_j=\frac{\rho^{2(j-1)}}{S_N}.
 \label{eq:onePortProfile}
\end{equation}
Its retention participation number, defined as the inverse quadratic
concentration of the normalized profile, is
\begin{equation}
 P_{\Ccal}\equiv
 \frac{\bigl(\sum_{j=1}^{N}\Ccal_j\bigr)^2}
 {\sum_{j=1}^{N}\Ccal_j^2}
 =\frac{1}{\sum_{j=1}^{N}\Ccal_j^2},
 \label{eq:retentionParticipation}
\end{equation}
where the last equality uses the one-port sum rule
\(\sum_j\Ccal_j=1\).
When $\rho>1$ and $N\gg1$,
\begin{align}
 \Ccal_N&\longrightarrow1-\rho^{-2},\label{eq:targetSaturation}\\
 \Ccal_1&\sim(\rho^2-1)\rho^{-2N},\label{eq:upstreamSuppression}\\
 P_{\Ccal}&\longrightarrow\frac{\rho^2+1}{\rho^2-1},\label{eq:onePortParticipation}
\end{align}
so the downstream-end mean-FI retention saturates at fixed \(U\).  The
upstream-end task is exponentially suppressed.  The
corresponding retention localization length is
\begin{equation}
 \xi_{\Ccal}=\frac{1}{2\ln\rho}.
 \label{eq:xiJordan}
\end{equation}
These statements concern normalized retention.  The absolute FI at fixed
internal resource also contains $\Ecal_j$, which may vary with position and
linewidth.

\suppsubsection{Source--internal-resource crossover in the unidirectional chain}
\label{sec:chainPareto}

For a single port at site 1, the preparation buildup is
\begin{equation}
 \beta_N=\norm{\bm X}^2=\frac{1}{\gamma^2}S_N.
 \label{eq:betaN}
\end{equation}
For \(\rho\ge1\), the weakest fixed-\(U\) task is the upstream
site, \(\Ccal_{\min}=1/S_N\).  The dimensionless preparation factor
\(\Gcal_N\equiv\gamma^2\beta_N=S_N\) obeys the one-port relation
\begin{equation}
 \Gcal_N\Ccal_{\min}=1.
 \label{eq:onePortGainCoverageProduct}
\end{equation}
This identity connects source buildup and worst-task retention for the uniform
geometric column.  With one input direction, the joint constraints are
\begin{equation}
 \bm s^\dagger\bm s\le P_0,
 \qquad
 \bm s^\dagger\bm H\bm s=\beta_N|s|^2\le U_0,
\end{equation}
so
\begin{equation}
 |s_\star|^2=\min(P_0,U_0/\beta_N).
\end{equation}
Relative to the full-state-drive FI at fixed \(U\) for task $j$,
\begin{equation}
\zeta_j(P_0,U_0)\equiv
\frac{F_{\mu,P,U,B,j}^{\star}}
{F_{\mu,U,\rm full,j}^{\star}}
=\Ccal_j\min\!\left[1,\frac{P_0}{U_0}\beta_N\right].
 \label{eq:jointOnePort}
\end{equation}
The second factor is the fraction of the internal-resource budget filled by
the available source power.
For the worst task at \(\rho\ge1\),
Eqs.~\eqref{eq:onePortGainCoverageProduct} and~\eqref{eq:jointOnePort} give
\begin{equation}
\zeta_{\min}(P_0,U_0)
 =\min\!\left(\frac{1}{S_N},\frac{P_0}{U_0\gamma^2}\right).
 \label{eq:jointWorstOnePort}
\end{equation}
The two terms are the geometric-retention ceiling and the source-limited
conversion factor, respectively.

For the skin-edge task $j=N$, the source-limited branch is
\begin{equation}
\zeta_N=\frac{P_0}{U_0}\frac{\rho^{2(N-1)}}{\gamma^2},
 \label{eq:targetSourceLimited}
\end{equation}
where the normalization sum cancels exactly.  Thus the source-limited branch
for the downstream task grows exponentially with \(N\) until it reaches the
fixed-\(U\) ceiling of Eq.~\eqref{eq:targetSaturation}.  The
upstream-end source-limited task is
\begin{equation}
\zeta_1=\frac{P_0}{U_0\gamma^2},
\end{equation}
independent of $N$ before the internal-resource constraint activates.  After
the crossover it follows Eq.~\eqref{eq:upstreamSuppression}.

The crossover condition is
\begin{equation}
 \frac{P_0}{U_0\gamma^2}S_N=1.
 \label{eq:exactCrossover}
\end{equation}
For $\rho>1$, treating $N$ continuously gives
\begin{equation}
 N_c\simeq
 \frac{\ln[1+(\rho^2-1)\gamma^2U_0/P_0]}{2\ln\rho},
 \label{eq:NcSupp}
\end{equation}
with an integer crossover obtained from the smallest $N$ satisfying
Eq.~\eqref{eq:exactCrossover}.  At $\rho=1$, the corresponding value is
$N_c=\gamma^2U_0/P_0$.  For $0<\rho<1$, $S_N$ saturates and the crossover may
remain absent.  Equation~\eqref{eq:targetSourceLimited} and the local-retention
bound quantify the normalized tradeoff.  At fixed downstream emission,
increasing \(\rho\) lowers the emission-normalized source cost of the downstream
task but reduces the upstream retention within each locally controlled interval.

For several ports, source power can be redistributed across the nested columns,
and the joint problem no longer reduces to one scalar.
Equation~\eqref{eq:jointDualSupp} then identifies the source-, mixed-, and
internal-resource-limited regions.

\suppsection{Lattice extensions and pole-identical counterexamples}
\suppsubsection{Hatano--Nelson bulk-retention crossover}
\label{sec:HN}

The open Hatano--Nelson generator~\cite{HatanoNelson1996,YaoWang2018,YokomizoMurakami2019,Okuma2020} is
\begin{equation}
 \bm H_g=-\ii\gamma\bm I+
 t\sum_{j=1}^{N-1}\left(e^g|j+1\rangle\langle j|+e^{-g}|j\rangle\langle j+1|\right).
 \label{eq:HNmodel}
\end{equation}
Let
\begin{equation}
 \bm D_g=\diag(1,e^g,e^{2g},\ldots,e^{(N-1)g}).
\end{equation}
Under open boundaries,
\begin{equation}
 \bm H_g=\bm D_g\bm H_0\bm D_g^{-1},
 \qquad
 \bm G_g(\omega)=\bm D_g\bm G_0(\omega)\bm D_g^{-1}.
 \label{eq:HNSimilarity}
\end{equation}
The open-boundary spectrum is unchanged by $g$.  The right and left response
envelopes are reweighted.

For a local port at $p$,
\begin{equation}
 (G_g)_{jp}=e^{g(j-p)}(G_0)_{jp}.
 \label{eq:HNGreen}
\end{equation}
Suppose the reciprocal Green function along the relevant direction has asymptotic envelope
\begin{equation}
 |(G_0)_{jp}|\sim e^{-\mu(\omega,\gamma)|j-p|}.
 \label{eq:reciprocalDecay}
\end{equation}
With full-state monitoring, white parameter-independent covariance,
\(\bm Q=\bm I\), one local source, and onsite tasks, \(\Ccal_j\) reduces
to the normalized squared Green-column amplitude.  Its bulk
finite-difference slope is
\begin{equation}
 \Delta_j\ln\Ccal_j\simeq2[g-\mu(\omega,\gamma)]
 \label{eq:HNExponentSupp}
\end{equation}
in the bulk; finite-chain reflections and normalization shape the boundary
layers.  The retention profile localizes downstream when
\begin{equation}
 g>\mu(\omega,\gamma).
 \label{eq:HNThresholdSupp}
\end{equation}
The profile remains source-side localized for \(g<\mu\).  This threshold
describes mean-FI retention relative to the ideal full-state drive benchmark.  Incident-flux NHSE
QFI uses a different resource comparison~\cite{McDonaldClerk2020,Bao2022}.

For the reciprocal uniform chain, the complex spatial equation is
\begin{equation}
 t(q+q^{-1})=\omega+\ii\gamma.
\end{equation}
The root with $|q|<1$ determines $\mu=-\ln|q|$.  In the downstream-localized region the retention localization length is
\begin{equation}
 \xi_{\Ccal}=[2(g-\mu)]^{-1}.
 \label{eq:HNxi}
\end{equation}

\begin{figure}[t]
\centering
\includegraphics[width=8.6cm]{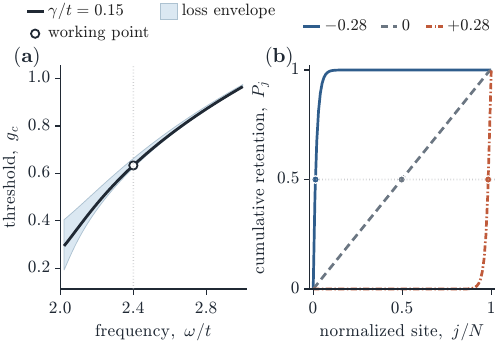}
\caption{\textbf{Hatano--Nelson bulk-retention crossover.}
 (a) The dimensionless asymptotic threshold
\(g_c(\omega,\gamma)=\mu(\omega,\gamma)\).  The dark curve uses
\(\gamma/t=0.15\), the shaded envelope spans
\(0.05\le\gamma/t\le0.30\), and the circle marks
\(\omega/t=2.4\), where \(g_c=0.6335\).
(b) Cumulative \(N=100\) task-specific mean-FI retention
\(P_j=\sum_{n\le j}\Ccal_n\) below, at, and above the crossover, for
\(\delta g\equiv g-g_c=-0.28,0,+0.28\).  The circles mark the sites at
which half of the one-port rank budget has accumulated.}
\label{fig:SHN}
\end{figure}

Figure~\ref{fig:SHN} shows $g_c=\mu$ and cumulative retention for
$N=100$, $\omega/t=2.4$, and $g-g_c=-0.28,0,0.28$.  The scan covers
$g\in[0,1.20]$.  Direct inversion and the similarity route differ by
\(\ValHNDirectSimilarity\); the direct finite-chain and fitted-profile
exponent errors are \(\ValHNDirectAsymptotic\) and \(\ValHNExponent\).

\suppsubsection{Non-Hermitian SSH lattice: separating topology and skin redistribution}
\label{sec:NHSSH}

For a two-sublattice test, we use an open non-Hermitian SSH chain
~\cite{Kunst2018,YaoWang2018,KochBudich2022,Sarkar2024Critical,KiorpelidisMakris2025}
with $N=2N_c$ sites,
\begin{align}
 \bm H_{\rm SSH}(g)={}&-\ii\gamma\bm I
 +\sum_{n=1}^{N_c}\left[t_1e^g|B_n\rangle\langle A_n|
 +t_1e^{-g}|A_n\rangle\langle B_n|\right]\nonumber\\
 &+\sum_{n=1}^{N_c-1}\left[t_2e^g|A_{n+1}\rangle\langle B_n|
 +t_2e^{-g}|B_n\rangle\langle A_{n+1}|\right].
 \label{eq:NHSSHmodel}
\end{align}
With the site ordering $(A_1,B_1,A_2,B_2,\ldots)$ and
\begin{equation}
 \bm D_g=\diag(1,e^g,e^{2g},\ldots,e^{(2N_c-1)g}),
\end{equation}
this model obeys the open-boundary similarity relation
\begin{equation}
 \bm H_{\rm SSH}(g)=\bm D_g\bm H_{\rm SSH}(0)\bm D_g^{-1}.
 \label{eq:NHSSHsimilarity}
\end{equation}
The open-boundary spectrum is identical for every $g$.

\begin{figure}[t]
\centering
\includegraphics[width=13.3cm]{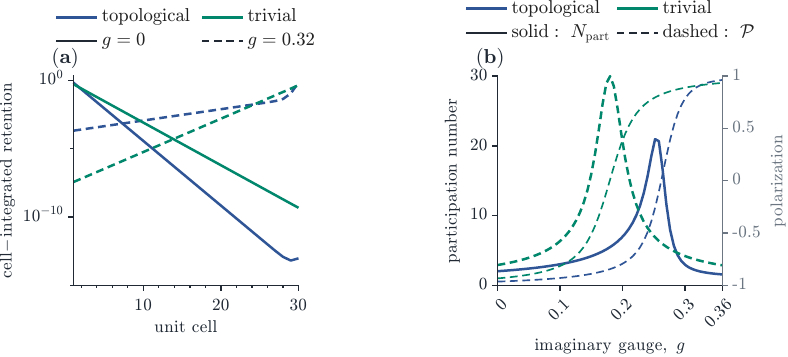}
\caption{\textbf{NHSSH spatial profiles.}  (a) Cell-summed mean-FI retention at
fixed $U$ for reciprocal and
gauge-transformed topological and trivial SSH chains.  Color identifies the
termination (blue, topological; green, trivial); solid and dashed curves denote
\(g=0\) and \(g=0.32\).  (b) The same colors identify the two terminations.
Solid curves and the left axis give \(N_{\rm part}\); dashed curves and the
gray right axis give the polarization \(\Pcal\), both defined in
Eq.~\eqref{eq:cellParticipationPolarization}.}
\label{fig:SNHSSH}
\end{figure}

We compare $t_1/t_2=0.6$ and $1.4$, which represent the reciprocal topological
and trivial regimes for the chosen termination
~\cite{SuSchriefferHeeger1979,ArandesBergholtz2025}.  The Gaussian
experiment uses $\bm C=\bm\Sigma_c=\bm Q=\bm I$, onsite tasks, and a port at
site 1, so each projector quotient is the physical-port/full-state-drive Fisher
ratio.  We take $N_c=30$, $\gamma/t_2=0.15$, $\omega=0$, and
$0\le g\le0.36$.  Cell summation removes sublattice staggering:
\begin{equation}
 \Ccal_n^{\rm cell}=\Ccal_{A_n}+\Ccal_{B_n}.
 \label{eq:cellCoverage}
\end{equation}
Because \(\sum_n\Ccal_n^{\rm cell}=1\) for the single-port calculation, we
characterize the cell profile by
\begin{equation}
 N_{\rm part}=\frac{1}{\sum_{n=1}^{N_c}(\Ccal_n^{\rm cell})^2},
 \qquad
 \Pcal=\sum_{n=1}^{N_c}x_n\Ccal_n^{\rm cell},
 \qquad
 x_n=-1+\frac{2(n-1)}{N_c-1}.
 \label{eq:cellParticipationPolarization}
\end{equation}
Thus \(N_{\rm part}\) is the cell-level analogue of
Eq.~\eqref{eq:retentionParticipation}.  The polarization \(\Pcal\in[-1,1]\)
records the center of the normalized retention profile.
Figure~\ref{fig:SNHSSH} separates the reciprocal edge envelope from its
imaginary-gauge redistribution.

With the frequency, loss, task, port, metric, and covariance fixed, the
imaginary gauge redistributes the reciprocal edge envelope without changing
the open-boundary spectrum.

\suppsection{Reduced-order circuit, covariance, and nuisance parameters}
\suppsubsection{Reduced-order noisy unilateral RLC/VCCS network}
\label{sec:RLC}

\suppsubsubsection{Network equations and energy metric}

The reduced nodal-admittance model is motivated by non-Hermitian
topolectrical and integrated-circuit sensors
~\cite{Yuan2023Circuit,Deng2024Circuit} and by unidirectional circuit coupling
based on a voltage follower~\cite{Zhao2024}.  For peak-amplitude phasors,
\begin{equation}
 \bm Y(\Omega)=\ii\Omega\bm C+\bm G_{\rm c}+(\ii\Omega\bm L)^{-1},
 \label{eq:Ymatrix}
\end{equation}
where $\bm C=\diag(C_j)$, $\bm L=\diag(L_j)$, and
\begin{equation}
 (G_{\rm c})_{jj}=R^{-1},
 \qquad
 (G_{\rm c})_{j+1,j}=-g_j.
 \label{eq:Gconductance}
\end{equation}
A coherent current drive $\bm i_{\rm in}=\bm B\bm s$ produces the mean
node-voltage phasor
\begin{equation}
 \bar{\bm v}=\bm Z\bm B\bm s,
 \qquad \bm Z=\bm Y^{-1}.
 \label{eq:networkResponse}
\end{equation}
The time-averaged electric and magnetic energy of this coherent mean field is
\begin{equation}
 U_{\rm coh}=\bar{\bm v}^{\dagger}\bm Q\bar{\bm v},
 \qquad
 \bm Q=\frac14\bm C+\frac{1}{4\Omega^2}\bm L^{-1}.
 \label{eq:RLCenergy}
\end{equation}
Equation~\eqref{eq:RLCenergy} constrains only the coherent mean field; thermal,
active, and readout fluctuations enter the likelihood.  The metric is
calibrated at $\epsilon=0$, separately for each disorder realization.  If RMS
phasors are used, the same physical energy is written with
\(\bm Q_{\rm RMS}=\bm C/2+\bm L^{-1}/(2\Omega^2)\).  This common rescaling of
the metric and energy budget leaves the retention ratios unchanged.

For uniform components at resonance $\Omega_0=(LC)^{-1/2}$, the diagonal
admittance is $R^{-1}$ and the impedance matrix is lower triangular.  A port at
node 1 generates the geometric voltage ratio
\begin{equation}
 \frac{|v_{j+1}|}{|v_j|}=g_m R\equiv\rho.
 \label{eq:RLCRho}
\end{equation}
Here \(\rho=g_m R\) is the forward-buildup ratio.  The circuit topology remains
strictly nonreciprocal for all \(\rho\); changing \(\rho\) changes only the
forward buildup.
Because $\bm Q$ is diagonal and uniform at the nominal point, the one-port,
task-specific mean-FI retention at fixed \(U_{\rm coh}\) for the local
capacitance task below is exactly the geometric law of
Eq.~\eqref{eq:onePortProfile}.

The parameter task is a fractional local capacitance change $C_j\mapsto C_j(1+\epsilon)$.  The admittance tangent is
\begin{equation}
 \partial_\epsilon\bm Y=\ii\Omega C_j|j\rangle\langle j|,
 \qquad
 \partial_\epsilon\bm Z=-\bm Z(\partial_\epsilon\bm Y)\bm Z.
 \label{eq:dZcap}
\end{equation}
The nominal simulations monitor the full noisy voltage vector.  Premultiplying
by an output selection matrix gives the corresponding partial-monitoring model.

\suppsubsubsection{Noise covariance}

The input-referred current-noise covariance is the sum of passive Johnson noise
and a phenomenological effective active-source contribution
~\cite{Johnson1928,Nyquist1928},
\begin{align}
 \bm S_I^{\rm pass}&=4k_{\rm B}T R^{-1}\bm I,\label{eq:JohnsonNetwork}\\
 \bm S_I^{\rm act}&=4k_{\rm B}T_{\rm act}\diag(0,g_1,\ldots,g_{N-1}).
 \label{eq:ActiveNetworkNoise}
\end{align}
The output voltage covariance is
\begin{equation}
 \bm\Sigma_c=\bm Z(\bm S_I^{\rm pass}+\bm S_I^{\rm act})\bm Z^\dagger
 +\sigma_r^2\bm R_c,
 \qquad
 (R_c)_{jk}=r^{|j-k|}.
 \label{eq:RLCcovariance}
\end{equation}
For $T=300$ K and $T_{\rm act}=600$ K, the proper complex record covariance is
defined by
\begin{equation}
 \mathbb E[\bm n\bm n^\dagger]
 =\bm S_{n}^{(1)}B_{\rm eq}\equiv\bm\Sigma_c,
 \qquad
 \mathbb E[\bm n\bm n^{\mathsf T}]=\bm0,
 \label{eq:oneSidedProperConvention}
\end{equation}
where $B_{\rm eq}=1$ Hz.  The readout density is
$\sigma_r^2=10^{-18}~{\rm V^2/Hz}$ with $r=0.35$.  This covariance is used for
both drive sets; $T_{\rm act}$ parametrizes the reduced active-source noise.

For the local rank-one capacitance task, the mean FI at fixed
\(U_{\rm coh}\) follows directly from
\begin{equation}
 \bm A_j=-\bm\Sigma_c^{-1/2}\bm Z(\partial_\epsilon\bm Y_j)\bm Q^{-1/2},
\end{equation}
so
\begin{equation}
 F_{\mu,U,{\rm full},j}^{\star}=2U_0\norm{\bm A_j}_2^2,
 \qquad
 F_{\mu,U,B,j}^{\star}=2U_0\norm{\bm A_j\bm\Pi_X}_2^2.
 \label{eq:RLCdirectFI}
\end{equation}
For this rank-one admittance tangent, the common emission factor cancels from
the physical-port/full-state-drive mean-FI ratio whenever the full-state-drive
information is nonzero.  The current-noise spectrum \(\bm S_I\)
and readout covariance are fixed for the local-capacitance task, so the
covariance derivative is
\begin{equation}
 \partial_\epsilon\bm\Sigma_c=(\partial_\epsilon\bm Z)\bm S_I\bm Z^\dagger+\bm Z\bm S_I(\partial_\epsilon\bm Z)^\dagger,
 \label{eq:dSigmaRLC}
\end{equation}
where $\bm S_I=\bm S_I^{\rm pass}+\bm S_I^{\rm act}$.  Parameter-dependent
source or readout noise adds
\(\bm Z(\partial_\epsilon\bm S_I)\bm Z^\dagger+
\partial_\epsilon\bm\Sigma_{\rm read}\).  The total FI for each drive
configuration is obtained by adding this common covariance-FI block to its
mean-response block.

Since \(\bm Z\) is invertible, full-state control is equivalent to independent
current drives at all nodes.  Write
\(F_{\mu,{\rm full},j}(U_0)=a_j U_0\).  Because the physical-port mean block is
\(\Ccal_j a_j U_0\) and the covariance block is common to both input-port sets,
\begin{equation}
 \Rcal_{{\rm tot},j}(U_0)=
 \frac{\Ccal_j a_j U_0+F_{\Sigma,j}}
 {a_j U_0+F_{\Sigma,j}}.
 \label{eq:RLCtotalBudget}
\end{equation}
The mean-response and covariance component retentions are \(\Ccal_j\) and
\(1\), respectively.  Main-text Fig.~4(c)
plots their contributions to the full-state-drive total,
\(\Ccal_j a_j U_0/(a_j U_0+F_{\Sigma,j})\) and
\(F_{\Sigma,j}/(a_j U_0+F_{\Sigma,j})\), whose sum is
\(\Rcal_{{\rm tot},j}\).
Thus $\Rcal_{{\rm tot},j}\to1$ as $U_0\to0$ when covariance motion is
nonzero, and $\Rcal_{{\rm tot},j}\to\Ccal_j$ when the coherent mean dominates.
The small-\(U_0\) limit is a fixed-bath covariance baseline, independent of the
vanishing coherent probe.  The physical-port mean and covariance blocks are
equal at
\begin{equation}
 U_{\times,j}=\frac{F_{\Sigma,j}}{\Ccal_j a_j}.
 \label{eq:RLCbudgetCrossover}
\end{equation}
At the skin edge and $U_0=1$ aJ, the mean and total retentions are
\(\RLCNominalEdge\) and \(\RLCNominalTotal\); the normalized mean and
covariance contributions are \(\RLCMeanContributionAtOneAJ\) and
\(\RLCCovarianceContributionAtOneAJ\).  Their crossover occurs at
$U_{\times}=\RLCMeanCovarianceCrossoverAJ$ aJ.  Across
$10^{-22}$--$10^{-14}$ J, the total retention changes from
\(\RLCTotalRetentionBudgetLow\) to \(\RLCTotalRetentionBudgetHigh\).

\suppsubsubsection{Stability of the reduced network}

The time-domain nodal equation is
\begin{equation}
 \bm C\ddot{\bm v}+\bm G_{\rm c}\dot{\bm v}+\bm L^{-1}\bm v=\dot{\bm i}_{\rm in}.
 \label{eq:RLCtime}
\end{equation}
Because $\bm C$ and $\bm L$ are diagonal and $\bm G_{\rm c}$ is lower
triangular, the quadratic matrix polynomial is triangular.  Its poles equal the
isolated parallel-RLC roots
\begin{equation}
 s_{j,\pm}=\frac{-R^{-1}\pm\sqrt{R^{-2}-4C_j/L_j}}{2C_j},
 \qquad
 \Gamma_j=-\max_{\pm}\operatorname{Re}s_{j,\pm}>0.
 \label{eq:RLCmargin}
\end{equation}
In the nominal underdamped regime,
$\Gamma_j=(2RC_j)^{-1}$ and $\min_j\Gamma_j=7.81\times10^6$ s$^{-1}$.
This stability analysis applies to the reduced VCCS model.  Device-level
dynamics add bandwidth, internal poles, saturation, and loading.

The nominal component values are
\begin{equation}
 N=40,
 \quad C=3.2~{\rm nF},
 \quad L=20~{\rm nH},
 \quad R=20~\Omega,
 \quad g_m=0.0675~{\rm S},
 \label{eq:RLCparams}
\end{equation}
which give
\begin{equation}
 f_0=\frac{1}{2\pi\sqrt{LC}}=19.894~{\rm MHz},
 \qquad \rho=g_m R=1.35.
\end{equation}
The nominal downstream and upstream mean-FI retentions are $\RLCNominalEdge$
and $\RLCNominalUpstream$, respectively.

\suppsubsubsection{Disorder and frequency selection}

\begin{figure}[t]
\centering
\includegraphics[width=8.6cm]{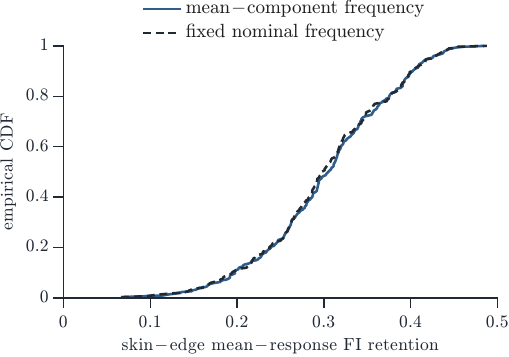}
\caption{\textbf{Network disorder.}  Empirical distributions of skin-edge
mean-FI retention for \(\RLCSamples\) realizations with 5\% component
disorder.  The blue solid curve uses the sample's mean-component resonance
estimate \(f_{\rm mc}\); the charcoal dashed curve uses the fixed nominal
frequency.  Their maximum empirical-CDF separation is
\(\RLCFrequencyProtocolKSDistance\).}
\label{fig:SRLCdisorder}
\end{figure}

For each Monte Carlo sample, independent Gaussian perturbations of standard
deviation 5\% are applied to every $C_j$, $L_j$, and $g_j$.  The aggregate
mean-component resonance estimate
\(f_{\rm mc}=[2\pi\sqrt{\overline C\,\overline L}]^{-1}\), with bars denoting
sample means, is used for all tasks in that realization.  Results at the fixed
nominal frequency are recorded separately.  All 300 samples retain positive
components and the feed-forward reduced model remains stable.  At
\(f_{\rm mc}\), the downstream mean-FI retention has median
$\RLCMedian$, fifth percentile $\RLCQFive$, and ninety-fifth percentile
$\RLCQNinetyFive$.  A fixed-seed, nonparametric bootstrap with
\(\RLCBootstrapReplicates\) resamples gives the respective 95\% confidence intervals
\([\RLCMedianCILow,\RLCMedianCIHigh]\),
\([\RLCQFiveCILow,\RLCQFiveCIHigh]\), and
\([\RLCQNinetyFiveCILow,\RLCQNinetyFiveCIHigh]\).
The smallest stability margin is $\RLCMinStability$ s$^{-1}$.
Figure~\ref{fig:SRLCdisorder} compares the two frequency choices over the full
empirical distribution.

The frequency estimate is sample specific and independent of the local task
position.  For an unknown task position, a common frequency,
multifrequency schedule, or prior-weighted allocation defines the acquisition
protocol.

\suppsubsection{Pole-identical PT--UC pair and diabolic-point reference}
\label{sec:benchmarks}

The loss-shifted parity--time-symmetric (PT) and unidirectional-coupling (UC)
dimers form a pole-identical pair; the diabolic-point (DP) dimer is a normal
reciprocal reference.  Their common pole trajectories isolate differences in
preparation, emission, and active-noise contributions to the Fisher operators
under the chosen resource.  Amplification noise, mode
nonorthogonality, linewidth, coherent absorption, and readout affect EP
measurements~\cite{Zhang2019,Xiao2019,Wiersig2020,
LoughlinSudhir2024,Wang2021CPA,Wang2026CPAEP3}; the covariance model below
specifies the corresponding passive and added-noise terms for the dimer
comparison.

\suppsubsubsection{Generators and matched pole splitting}

In dimensionless frequency units, the three internal generators are
\begin{align}
 \bm h_{\rm PT}(\epsilon)&=
 \begin{pmatrix}
 1-\epsilon-\ii\gamma+\ii g & g+\epsilon\\
 g+\epsilon & 1-\epsilon-\ii\gamma-\ii g
 \end{pmatrix},
 \label{eq:hPT}\\
 \bm h_{\rm UC}(\epsilon)&=
 \begin{pmatrix}
 1-\epsilon-\ii\gamma & \epsilon\\
 \kappa+\epsilon & 1-\epsilon-\ii\gamma
 \end{pmatrix},
 \label{eq:hUC}\\
 \bm h_{\rm DP}(\epsilon)&=
 \begin{pmatrix}
 1-\epsilon-\ii\gamma & \epsilon\\
 \epsilon & 1-\epsilon-\ii\gamma
 \end{pmatrix}.
 \label{eq:hDP}
\end{align}
The parameter tangent used in all three cases is
\begin{equation}
 \bm V=-\bm u\bm u^\dagger,
 \qquad
 \bm u=(1,-1)^{\mathsf T}.
\label{eq:benchmarkV}
\end{equation}
The shared rank-one tangent can result from eliminating an auxiliary mode
coupled to the resonators with opposite phases.  For \(\kappa=2g\), PT and UC
have identical complex poles throughout the reported perturbation interval
$10^{-5}\le\epsilon\le0.20$,
\begin{equation}
 \Delta\lambda_{\rm PT}=\Delta\lambda_{\rm UC}
 =2\sqrt{\epsilon(\kappa+\epsilon)}.
 \label{eq:matchedSplitting}
\end{equation}
The common radiative port rate \(\eta\) is included through
\begin{equation}
 \bm G=(\bm h-\ii\eta\bm I-\omega\bm I)^{-1},
 \qquad
 \bm K_\eta=\sqrt{2\eta}\,\bm I,
 \qquad
 \bm S=-\bm I-\ii\bm K_\eta\bm G\bm K_\eta.
 \label{eq:benchmarkScattering}
\end{equation}
On the reported interval the common square root is real and the radiative
poles have imaginary part $-(\gamma+\eta)$.  The nominal parameters are
\begin{equation}
 \eta=0.025,\quad \kappa=0.312,\quad g=0.156,\quad
 \gamma=0.010,\quad N_0=1,\quad N_{\rm read}=0,
 \label{eq:benchmarkNominal}
\end{equation}
At the reference point we set $\epsilon=10^{-3}$, optimize over
$0.35\le\omega\le1.45$, use both input and output ports, and take
$\bm R=\bm Q=\bm I$.  This equal-modal metric excludes
dc pump power, saturation, and gain-bandwidth costs.

\suppsubsubsection{Power-wave covariance and fluctuation--dissipation closure}

All covariance matrices are expressed in one power-wave normalization~\cite{Caves1982,Clerk2010}.  The passive and added bath terms are
\begin{align}
 \bm\Gamma_{{\rm p},{\rm PT}}&=\gamma\bm I+g\operatorname{diag}(0,1),
 &\bm\Gamma_{{\rm add},{\rm PT}}&=g\nu_{\rm PT}\operatorname{diag}(1,0),\nonumber\\
 \bm\Gamma_{{\rm p},{\rm UC}}&=\gamma\bm I,
 &\bm\Gamma_{{\rm add},{\rm UC}}&=\kappa\nu_{\rm UC}\operatorname{diag}(0,1),\nonumber\\
 \bm\Gamma_{{\rm p},{\rm DP}}&=\gamma\bm I,
 &\bm\Gamma_{{\rm add},{\rm DP}}&=\bm0,
 \label{eq:benchmarkBaths}
\end{align}
The marked reference point uses \(\nu_{\rm PT}=1\) and
\(\nu_{\rm UC}=g/\kappa=1/2\), so
\(\Tr\bm\Gamma_{{\rm add},{\rm PT}}
=\Tr\bm\Gamma_{{\rm add},{\rm UC}}\).
For a passive bath at the incident-channel reference temperature,
\begin{equation}
 \bm\Sigma_c=N_0\bm S\bm S^\dagger
 +4\eta N_0\bm G\bm\Gamma_{\rm p}\bm G^\dagger
 +\bm\Sigma_{\rm add}.
 \label{eq:benchmarkSigma}
\end{equation}
For dynamically consistent passive isothermal baths and
\(\bm\Sigma_{\rm add}=0\),
\begin{equation}
 \bm S\bm S^\dagger
 +4\eta\bm G\bm\Gamma_{\rm p}\bm G^\dagger=\bm I,
 \qquad
 \bm\Sigma_c=N_0\bm I.
 \label{eq:FDTclosure}
\end{equation}
The passive DP reference satisfies Eq.~\eqref{eq:FDTclosure} with residual
$\ValFDTResidual$.  This power-wave convention includes the impedance
conversion from a Thevenin voltage-noise density.

For architecture $A\in\{\mathrm{PT},\mathrm{UC},\mathrm{DP}\}$, the additional
covariance uses the matrix defined in Eq.~\eqref{eq:benchmarkBaths}:
\begin{equation}
 \bm\Sigma_{{\rm add},A}=4\eta N_0\bm G_A
 \bm\Gamma_{{\rm add},A}
 \bm G_A^\dagger+N_{\rm read}\bm I.
 \label{eq:activeNoiseBenchmark}
\end{equation}
The multipliers \(\nu_{\rm PT}\) and \(\nu_{\rm UC}\) are scanned independently
within this reduced small-signal covariance model.

\suppsubsubsection{Results under source and internal constraints}

At the common reference point, the preparation Gramian enters the target-only mean-FI
operator through the spectral filter of Eq.~\eqref{eq:WtauIdentity}.  Define the
normalized coefficients
\begin{equation}
 \phi_P=\frac{F_{\mu,P}^\star}{2P_0}
 =\lambda_{\max}(\bm A\bm W_0\bm A^\dagger),
 \qquad
 \phi_U=\frac{F_{\mu,U}^\star}{2U_0}
 =\lambda_{\max}(\bm A\bm\Pi_X\bm A^\dagger).
 \label{eq:benchmarkFilterRates}
\end{equation}
PT and UC have the same two eigenvalues of \(\bm W_0\), but their
\(\phi_P\) and \(\phi_U\) differ because \(\bm W_0\) has a different
orientation relative to \(\bm A^\dagger\bm A\).  At the fixed-\(U\) endpoint,
both dimers have full reachable rank; their
remaining difference is set by parameter action and noise-whitened emission.

\begin{figure}[t]
\centering
\includegraphics[width=8.6cm]{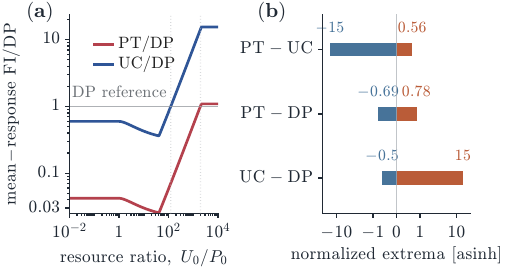}
\caption{\textbf{Identical poles do not determine a Fisher ordering.}
 (a) Joint-resource mean-FI optima for PT and UC relative to the diabolic-point reference
versus \(U_0/P_0\), under the stated calibration.  The corners mark changes
of the active resource constraint.  (b) Minimum and maximum eigenvalues of
the three pairwise resource-whitened Fisher-operator differences at
\((\epsilon,\omega)=(10^{-3},1)\).  Every interval contains both signs,
 ruling out a uniform ordering over all admissible probes at this point.  The horizontal axis uses an
inverse-hyperbolic-sine display coordinate
\(x=\operatorname{asinh}(\lambda/0.5)\); ticks at
\(\lambda=0,\pm1,\pm10\) and the endpoint labels report the original
DP-normalized eigenvalues.}
\label{fig:SpoleOrdering}
\end{figure}

At \(\epsilon=10^{-3}\), optimizing frequency over the common grid gives the
following values.  The columns \(\beta_P\) and \(\Ecal_P\) are evaluated at the
fixed-source optimum and denote
\(\beta_P=\norm{\bm X^\dagger\bm w}^2/
(\bm w^\dagger\bm\Pi_X\bm w)\) and
\(\Ecal_P=\norm{\bm\Sigma_c^{-1/2}\bm C\bm G\bm v}^2\), respectively.
\begin{center}
\begin{tabular}{lrrrrrr}
\toprule
Model & \(F_{\mu,P}^\star\) & \(F_{\mu,U}^\star\) & \(\omega_P^\star\) & \(\omega_U^\star\) & \(\beta_P\) & \(\Ecal_P\)\\
\midrule
PT & \(1.5204\times10^4\) & \(18.338\) & \(0.99207\) & \(0.89488\) & \(1067.8\) & \(3.5597\)\\
UC & \(2.1505\times10^5\) & \(205.46\) & \(0.99269\) & \(0.98012\) & \(1070.5\) & \(50.222\)\\
DP & \(1.3328\times10^4\) & \(326.53\) & \(0.99800\) & \(0.99800\) & \(40.816\) & \(81.633\)\\
\bottomrule
\end{tabular}
\end{center}
The fixed-\(U\) column is optimized at its own frequency.  Absolute Fisher
coefficients are rounded to five significant digits; all entries use the
equal-added-bath-trace calibration.
Figure~\ref{fig:SpoleOrdering}(a) shows the resulting joint-resource
frontiers, while Fig.~\ref{fig:SpoleOrdering}(b) tests pairwise operator
ordering at the common reference point.

\suppsubsection{Covariance information and nuisance parameters}
\label{sec:covNuisance}

\suppsubsubsection{Covariance Fisher contribution}

For the Gaussian model of Sec.~\ref{sec:gaussianFI},
\begin{equation}
 F_{\rm tot}=F_\mu+F_\Sigma,
 \qquad
 F_\Sigma=\Tr[(\bm\Sigma_c^{-1}\partial_\epsilon\bm\Sigma_c)^2].
 \label{eq:totalFI}
\end{equation}
The projector theorem constrains $F_\mu$.  Covariance FI is a per-record score
that adds to each drive configuration, as illustrated by the RLC example.

\suppsubsubsection{Nuisance-parameter elimination}

Let \(\bm\vartheta=(\epsilon,\bm\nu)\) contain the target and nuisance
parameters~\cite{Naikoo2023Multiparameter,Verreussel2026}.  Partition
\begin{equation}
 \bm F=
 \begin{pmatrix}
 F_{\epsilon\epsilon} & \bm F_{\epsilon\nu}\\
 \bm F_{\nu\epsilon} & \bm F_{\nu\nu}
 \end{pmatrix}.
\end{equation}
With a Gaussian calibration prior of precision \(\bm\Lambda_\nu\succeq0\), the
effective information for \(\epsilon\) is the generalized Schur complement
~\cite{Kay1993,VanTrees2001}
\begin{equation}
 F_{\epsilon|\nu}=F_{\epsilon\epsilon}
 -\bm F_{\epsilon\nu}(\bm F_{\nu\nu}+\bm\Lambda_\nu)^+
 \bm F_{\nu\epsilon}.
 \label{eq:SchurNuisance}
\end{equation}
For a singular nuisance-plus-prior block, Eq.~\eqref{eq:SchurNuisance} requires
\begin{equation}
 \ran\bm F_{\nu\epsilon}\subseteq
 \ran(\bm F_{\nu\nu}+\bm\Lambda_\nu).
 \label{eq:nuisanceSupport}
\end{equation}
The dimer nuisance vector contains resonance frequency, radiative coupling,
passive loss, complex input gain, and the architecture's noise scale, with
common fractional prior widths.  Repeated records scale both data blocks;
the external calibration prior is added once.

\suppsubsubsection{Common-probe budget}

For the orthonormal task basis of Sec.~\ref{sec:geometry}, let
\(\bm z\in\ran\bm X\) be an internal state with unit internal resource.  Its
common-probe overlaps are
\begin{equation}
 c_j^{\rm common}=|\widetilde{\bm w}_j^\dagger\bm z|^2,
 \qquad
 \sum_j c_j^{\rm common}=1.
 \label{eq:commonProbeBudget}
\end{equation}
For a rank-one task with whitened-emission factor \(\Ecal_j\), the corresponding
mean FI at internal resource \(U\) is
\(2U\Ecal_j c_j^{\rm common}\).  Here \(c_j^{\rm common}\) is the preparation
overlap multiplying the task's Fisher coefficient.

For the feed-forward chain, expand a common state in the disjoint segment
basis of Eq.~\eqref{eq:segmentBasis},
\begin{equation}
 |\bm z\rangle=\sum_{\ell=1}^{m}a_\ell|\bm u_\ell\rangle,
 \qquad \sum_{\ell=1}^{m}|a_\ell|^2=1.
 \label{eq:commonSegmentExpansion}
\end{equation}
Let
\begin{equation}
 d_\ell=\min_{p_\ell\le j<p_{\ell+1}}\Ccal_j.
 \label{eq:segmentMinimumDefinition}
\end{equation}
The weakest normalized overlap in segment \(\ell\) is
\(|a_\ell|^2d_\ell\).  Maximizing a common floor \(t\) therefore requires
\(|a_\ell|^2\ge t/d_\ell\) for every segment.  The normalization in
Eq.~\eqref{eq:commonSegmentExpansion} gives the attainable optimum
\begin{equation}
 \Ccal_{\rm common}^{\star}
 =\left(\sum_{\ell=1}^{m}d_\ell^{-1}\right)^{-1},
 \qquad
 |a_\ell|^2=
 \frac{d_\ell^{-1}}{\sum_{r=1}^{m}d_r^{-1}}.
 \label{eq:commonProbeExactSupp}
\end{equation}
Task-specific retuning instead gives
\begin{equation}
 \Ccal_{\rm task}^{\star}=\min_\ell d_\ell.
 \label{eq:taskProbeExactSupp}
\end{equation}
When all segments have equal length, all \(d_\ell\) coincide and
\(\Ccal_{\rm common}^{\star}=\Ccal_{\rm task}^{\star}/m\).  These are
retentions for individual tasks relative to the ideal full-state drive benchmark; unequal emission factors
weight the corresponding absolute Fisher max--min problem.

For an orthonormal ensemble \(\{\bm z_r\}_{r=1}^m\) spanning \(\ran\bm X\),
\begin{equation}
 \sum_{r=1}^m|\widetilde{\bm w}_j^\dagger\bm z_r|^2=\Ccal_j.
 \label{eq:ensembleCoverage}
\end{equation}
With unit internal resource per basis probe, Eq.~\eqref{eq:ensembleCoverage} uses
\(m\) times the internal and acquisition resource of a single-probe experiment.
Under a total budget \(U_{\rm tot}\), equal allocation yields
\((U_{\rm tot}/m)\Ccal_j\).  More generally, a common-probe experiment optimizes
a chosen scalar risk functional under its prior and total record budget
~\cite{ChalonerVerdinelli1995,Verreussel2026}.

\suppsection{Numerical methods and convergence}
\label{sec:numerics}

\suppsubsection{Reachable subspaces and covariance whitening}

The projector is evaluated using a numerical rank determined by a relative
tolerance after equilibrating the active column norms.  This separates the orientation of the
reachable subspace from the physical gain of each drive channel.  With
\(c_a=\norm{\bm X_{:a}}_2\) and \(c_{\max}=\max_a c_a\), column \(a\) enters the
range calculation when
\begin{equation}
 c_a>\tau_{\rm svd}\max(N,p)c_{\max}
 \label{eq:SVDcolumnThreshold}
\end{equation}
and the unit-norm columns form \(\overline{\bm X}\).  If
\(\overline{\bm X}=\bm U\overline{\bm\Sigma}\bm V^\dagger\), its retained
singular vectors satisfy
\begin{equation}
 \overline\sigma_k>
 \tau_{\rm svd}\max(N,p)\overline\sigma_1,
 \qquad \bm\Pi_X=\bm U_r\bm U_r^\dagger .
 \label{eq:SVDthreshold}
\end{equation}
We use \(\tau_{\rm svd}=10^{-12}\).  Balancing is used only to construct
\(\bm\Pi_X\); source costs and preparation singular values are computed from
the original map.  Positive-semidefinite square roots are evaluated in a
symmetrized Hermitian eigenbasis after nondimensionalization.  For
\(\bm R=\bm L^\dagger\bm L\), the amplitude threshold is
\(d_k>(n\tau_{\rm svd})^2d_{\max}\).  The analytic segment basis is used for
the feed-forward chain.  Sweeping \(\tau_{\rm svd}\) from \(10^{-14}\) to
\(10^{-8}\) leaves the rank unchanged in the models reported here; the largest
projector displacement from the \(10^{-14}\) reference is
\(\ValRankContractProjector\).

The reduced network covariance is positive definite and is whitened by a
Cholesky factor.  Its condition number is
\(\RLCNominalCovarianceCondition\), with relative spectral margin
\(\RLCNominalCovarianceRelativeMargin\).  Alternative fixed output
coordinates change the absolute RLC Fisher values by
\(\ValRLCCoordinateInvariance\) in double precision.  Dimensional Fisher
values are therefore quoted to at most five significant digits.  The
dimensionless retention and mean--covariance crossover are stable at the
reported precision.

For a strongly conditioned covariance, a sufficiently small covariance
derivative cannot be separated from double-precision whitening error.  We
characterize this numerical scale by
\begin{equation}
 F_{\Sigma,\mathrm{num}}(\eta)
 =q\left[\eta\epsilon_{\mathrm{mach}}
 \kappa_2(\bm\Sigma_c)\right]^2,
 \qquad \eta\in\{10,20,50\}.
 \label{eq:FcovResolutionFloor}
\end{equation}
This scale sets the numerical resolution.  The skin-edge task and its
mean--covariance crossover remain above it for \(\eta=10,20,50\).

\suppsubsection{Optimization and convergence}

The joint-resource frontier is obtained from the primal--dual form of
Eq.~\eqref{eq:jointDualSupp}; the three branches satisfy the corresponding KKT
relations.

Hatano--Nelson slopes are fitted away from both boundaries.  For NHSSH, the
large-$g$ similarity matrix is ill conditioned, so we compare
\begin{equation}
 \delta_{\rm sim}=
 \frac{\norm{\bm H_g-\bm D_g\bm H_0\bm D_g^{-1}}_{\rm F}}
 {\norm{\bm H_g}_{\rm F}},
 \label{eq:similarityResidual}
\end{equation}
whose maximum discrepancy is \(\ValNHSSHSimilarity\).  Dimer frequency grids
are refined from 101 to 3201 points.  The random-matrix, dimer, RLC-disorder,
and RLC-bootstrap seeds are 20260720, 20260721, 20260719, and 20260722,
respectively.  Table~\ref{tab:residuals} summarizes the numerical agreement of
the identities used in the article.

\begin{table}[!htbp]
\caption{Numerical agreement for identities used in the article.}
\label{tab:residuals}
\centering\small
\begin{tabular}{@{}p{11.8cm}r@{}}
\toprule
Calculation & maximum discrepancy\\
\midrule
Joint-resource primal--dual comparison & \(\ValJointGap\)\\
Efficient-score distance/Schur identity & \(\ValScoreSchur\)\\
Fixed-\(U\) projector identity, condition number $\le10^4$ & \(\ValFixedEnergyResolved\)\\
Local-port analytic/projector comparison & \(\ValLocalPort\)\\
RLC mean-response FI/projector retention & \(\ValRLCProjector\)\\
Hatano--Nelson fitted exponent & \(\ValHNExponent\)\\
NHSSH direct/similarity resolvent comparison & \(\ValNHSSHField\)\\
Passive power-wave FDT & \(\ValFDTResidual\)\\
\bottomrule
\end{tabular}
\end{table}

\suppsection{Extensions of the statistical model}
\label{sec:limitations}

Probe-dependent noise, nonlinear response, and parameter-dependent ports or
baths lead to the corresponding full-likelihood optimization.  Improper
complex Gaussian records use an augmented covariance containing the
pseudocovariance.  If the internal metric \(\bm Q\) is semidefinite, a
zero-\(\bm Q\)-cost direction that carries target information remains in the
support analysis; the fixed-\(U\) problem is unbounded unless another active resource
constrains that direction.  In a joint-resource problem, only directions in
the common zero-cost kernel of all active resource matrices that are also
information-null may be quotiented out.  Projectors and pseudoinverses are then
formed on the remaining numerical support.

For coherent Gaussian probes in a physically realizable bosonic input--output
embedding, including the auxiliary noise channels required to preserve the
commutation relations, a measurement that attains the displacement QFI yields
the same quadratic mean block.  Non-Gaussian probes and measurements require
the broader quantum formulation of
Refs.~\mbox{\cite{BraunsteinCaves1994,Ding2023,YokomizoClerkAshida2026}}.
Disordered, bidirectional, and higher-dimensional local networks lead to
related max--min design problems beyond the closed-form feed-forward chain.

\def\bibinfo#1#2{%
  \ifstrequal{#1}{title}{\textcolor{black}{#2}}{#2}}
\hypersetup{urlcolor=blue}
\bibliography{refs}

%% file: generated_numbers.tex

\providecommand{\RLCMedian}{0.306}
\providecommand{\RLCQFive}{0.166}
\providecommand{\RLCQNinetyFive}{0.421}

\providecommand{\RLCNominalEdge}{0.45130}
\providecommand{\RLCNominalTotal}{0.98745}
\providecommand{\RLCMeanCovarianceCrossoverAJ}{94.7}